\documentclass[useAMS,usenatbib,usegraphicx,mathabx]{mn2e}

\usepackage{lscape}
\title[Star-forming FIR galaxies and effects of confusion]
{A far-infrared survey at the North Galactic Pole I:
Nearby star-forming galaxies and effect of confused sources on source counts}
\author[P. V\"ais\"anen et al.]
       {Petri V\"ais\"anen$^{1,2}$\thanks{petri@saao.ac.za}, 
          Jari K. Kotilainen$^{3}$,
          Mika Juvela$^{4}$, Kalevi Mattila$^{4}$, \newauthor
          Andreas Efstathiou$^{5}$,
          Jere Kahanp\"a\"a$^{4}$ \\
          $^{1}$South African Astronomical Observatory, P.O.Box 9,
                Observatory, 7935, Cape Town, South Africa\\
          $^{2}$Southern African Large Telescope Foundation, P.O.Box 9,
                Observatory, 7935, Cape Town, South Africa\\
          $^{3}$Tuorla Observatory, Department of Physics and Astronomy,
                University of Turku, V\"ais\"al\"antie 20,
                FI-21500 Piikki\"o, Finland \\
          $^{4}$Observatory,  P. O. Box
   14, FI-00014 University of Helsinki, Finland \\
          $^{5}$School of Sciences, European University Cyprus, 
                Diogenes Street, Engomi, 1516 Nicosia, Cyprus
}

\date{Accepted ...; Received ...; in original form .. .. }

\pagerange{\pageref{firstpage}--\pageref{lastpage}} \pubyear{2009}

\begin{document}

\maketitle

\label{firstpage}

\begin{abstract}

We present follow-up observations of the far-infrared (FIR) 
sources at 90, 150 and 180 $\umu$m detected as part of the ISOPHOT 
EBL project, which has recently measured the absolute surface brightness of 
the cosmic infrared background radiation (CIRB) for the first time 
independently from COBE data. We have observed the fields at
the North Galactic Pole region in the optical and near-IR, and complement
these data with SDSS photometry, and spectroscopy where available,
and present identifications of the 25 FIR sources which reach down to
$\sim150$ mJy in all three ISOPHOT bands. 
Identifications are done by means of full spectral energy density fitting
to all sources in the FIR error circle areas.  Approximately
80 per cent are identified as star-forming or star-bursting galaxies
at $z<0.3$.  We also find that more than half of the counterparts have
disturbed morphologies, with signs of past or present interactions.
However, only 20 per cent of all the sources are uniquely matched with a 
{\em single} galaxy -- 40 per cent are blends
of two or more of these nearby star-forming galaxies, while another 20
per cent are likely blends of nearby and fainter galaxies. The final
20 per cent are likely to be more luminous IR galaxies at higher redshifts.  
The blended sources have an effect on the FIR source counts.
In particular, taking into account realistic confusion or blending
of sources, the differential FIR counts move down by a factor of $\sim1.5$
and steepen in the 100 to 400 mJy range.

\end{abstract}


\section{Introduction}
\label{intro}

The initial detection of the 
cosmic infrared background (CIRB) in the far-infrared 
(FIR) using {\em COBE} between 100 and 240 $\umu$m 
by \citet{hauser98}, \citet{fixsen98}, and preliminarily by \citet{puget96},
showed that a large
part of all the radiation produced in the universe since the recombination 
epoch is contained in the CIRB.  In the context of understanding 
galaxy formation and evolution, and indeed the history of all luminous matter 
in the universe, it is thus important to understand what and where the 
CIRB contributors are. This has been a subject of intense work ever since the 
COBE results \citep[for reviews see][]{hauser01,lagache05}.

The ISOPHOT EBL project aimed to
determine the level of CIRB independently from the COBE data by making use 
of the better resolution of {\em ISO} which allows looking into the 
very deepest regions on the FIR sky in-between Galactic cirrus clouds. 
The FIR source counts from this project were reported in \citet{juvela00}.
The CIRB surface brightness in three wavelengths 90, 150 and 180 $\umu$m
was recently presented in \citet{juvela09} -- this constituted the first
independent verification of the absolute CIRB level since the various 
COBE results.
In between 150 and 180 $\umu$m the CIRB was measured to be $1.1\pm0.3$ 
MJy sr$^{-1}$ and at 90 $\umu$m a 2$\sigma$ upper limit of 2.3 MJy sr$^{-1}$ 
was measured, all values
consistent with the COBE measurements obtained with the DIRBE and FIRAS
instruments.

However, only the bright tail of FIR sources contributing to the 
CIRB can be resolved with data from past missions, reaching to
typical levels of somewhat fainter than $\sim100$ mJy at 180~$\mu$m.
The FIR source counts at 90 to 180~$\mu$m in our project 
contribute $\sim10-20$ 
per cent of the CIRB surface brightness \citep{juvela00}. 
This result is similar to other FIR studies 
\citep[e.g.][]{dole01,frayer06b,matsuura07} 
when looking at the FIR sources directly 
using {\em ISO}, {\em Spitzer}, and {\em AKARI}. 
Statistically one can probe fainter FIR populations by using e.g.\ stacking 
methods, as shown by \citet{dole06} and \citet{papovich07}.  
Due to significantly less severe confusion, 
the resolved fraction of individual sources in the mid-IR 15 and 24 $\umu$m
mid-IR wavelengths is much higher, around 60 to 75 per cent  
\citep[e.g.][]{elbaz02,papovich04,rodighiero06}. 
\citet{frayer06} report resolving 60 per cent of 
the CIRB also at 70~$\umu$m using fluctuation studies and 
extrapolated source counts.  At wavelengths longer than this, up 
until $\sim200$~$\umu$m, the fraction of {\em directly} 
resolved CIRB sources has essentially not changed since the various 
ISOPHOT studies. The recent
BLAST experiment reports resolving the FIR background at 500 $\umu$m by 
stacking a Spitzer 24 $\umu$m catalog against their FIR map
\citep{devlin09}.

During the decade of FIR follow-up studies it has also 
become evident how difficult it is to identify
the optical and near-IR counterparts of these FIR sources 
\citep[e.g.][]{patris03,taylor05,oyabu05,dennefeld05,
sajina06,frayer06b}. 
Typically only a third, or a half, of the FIR 
sources have unambiguous counterparts at shorter wavelengths.
The problem is this: because of the large beam sizes 
there are many possible counterparts in optical images 
within the positional error circles of FIR sources. 
Hence, to aid 
identification other waveband data, especially MIR or
radio data, and often quite elaborate decision trees are typically used
to merge catalogs with
unique counterparts \citep[e.g.][]{taylor05,oyabu05}. 
Another difficulty arising from the 
large beam sizes is the high level of confusion among the FIR
detections resulting
in blended objects, which in turn make the optical identifications even more
complicated. 
These difficulties indirectly affect mid-IR studies as well, since many 
interesting cosmological or galaxy evolution results from the MIR surveys 
require information of the spectral energy distributions (SEDs) of the FIR
part of the spectra of the targets. These are just not available to
similar depths, unless the MIR data itself is used to identify sources and
construct FIR SEDs, leading into possibly unhealthy dependencies.

Nevertheless, a basic expectation and consensus has 
been built 
where the FIR sources are either local 
IR-bright sources dominated by cold cirrus emission 
or higher redshift sources whose strong IR-luminosity 
raises them into the class of luminous or ultra-luminous IR-galaxies 
(LIRGs or ULIRGs,$>10^{11}{\rm L}_{\odot}$ or  $>10^{12}{\rm L}_{\odot}$,
respectively). 
The relative amounts of sources in these classes
depend on the depth of the survey \citep[see e.g.][]{rowan09}.  
To interpret the cosmological evolution of IR-bright galaxies, it
remains crucial to understand the exact nature of galaxies emitting at 
100 to 200 $\umu$m where the CIRB peaks.
The imminent Herschel data will probe these galaxies to 
further distances \citep[][]{pilbratt} and it is important 
to determine the 
characteristics of the more nearby bright FIR galaxy population for reference.  

In this paper we revisit the source detections of the ISOPHOT EBL 
project by presenting optical and NIR follow-up observations of the 
ISOPHOT NGP fields.  
Rather than trying to 
{\em force} individual optical and NIR counterparts for each FIR source, 
we assess carefully the possibility of confused, or blended, sources. 
In addition, we do not use mid-IR or radio data to identify sources.
We present the properties of all the unambiguous matches and discuss the
level of likely confusion in the FIR source list.
If it were to be shown that a significant fraction of ISOPHOT sources are 
actually confused sums of several objects, or even whole galaxy groups 
or clusters, FIR source counts and their evolutionary 
interpretation might have to be revised.  

In a second paper of the series (V\"ais\"anen et al., in preparation)
we will concentrate on the correlations of the FIR source positions with 
significant peaks in the galaxy surface density, any signs of 
clustering and any {\em bona fide} galaxy clusters.

We use a spatially flat cosmology with $H_{0}=71$ km s$^{-1}$ 
Mpc$^{-1}$, $\Omega=1$, and $\Omega_{m}=0.27$ throughout.

\section{Data}
\label{data}

\subsection{Observations and data reductions}

The ISOPHOT observations are fully described in \citet{juvela00}. 
The North Galactic Pole (NGP) field with an area of 1.07 sq.deg 
covers 2/3 of the whole ISOPHOT EBL project area and contains 
22 individual sources in the 170 and 150  $\umu$m bands, and 25 sources 
at 90 $\mu$m. Sources were
detected down to $\sim150$ mJy in all three bands, while simulations indicated
a typical 70 per cent or more completeness at $\sim200$ mJy. 
The three 'extra' detections in the 90 $\umu$m are in fact 
sources very close to another 90 $\umu$m source, and the two of them
share the same single longer wavelength ISOPHOT detection; we therefore
have followup of 22 different fields-of-view.  
All these fields were targeted with ground based observations presented in 
this paper. Observations were performed using the 2.56-m Nordic
Optical Telescope (NOT) on La Palma.  Optical data were taken
on 8-10 April 2003 using the ALFOSC instrument, and the near-IR observations
were carried out on 13-15 April 2003 using the NOTCam instrument.

ALFOSC is a 2x2k CCD camera, giving a field of view of 6.4'x6.4'. 
We obtained 900 sec $R$ and $I$-band (filters 76 and 12) images of the target 
fields, split into three separate offset integrations.  The data were 
reduced and combined in the standard manner using IRAF. 

NOTCam is a 1x1k ``Hawaii'' near infrared camera.  
The low resolution mode of the instrument was used,
giving a 4'x4' field of view.  We obtained $K$-band images of all the fields.
Target fields were observed in 
1-minute dithers for a total of 25-minute integration time per field.  
The flatfielding (twilight flats were used), 
sky frame construction, sky subtraction, and registering and
co-adding of the frames into final NIR images was performed using 
the XDIMSUM package along with dedicated IDL routines.

Seeing remained in the range $0.8 - 1.0$ arcsec in $K$-band and $0.9 - 1.3$ 
in the optical bands throughout our observing runs.
The astrometry for both optical and infrared images was obtained 
through matching GSC2 stars -- we estimate the accuracy to be better than 
0.4\arcsec\ throughout the frames.

We note that while the NGP area has not been observed by other targeted
FIR surveys in the past, it will shortly be covered by the Herschel Thousand 
Degree Survey (PI:Eales.\footnote{http://herschel.esac.esa.int/Docs/KPOT/KPOT\_accepted.html\\\#KPOT\_seales01\_2})

\subsection{Photometry}
\label{photo}

All photometry was derived 
using the SExtractor (v2.3.2) software \citep{bertin}.
Both aperture and 'auto' magnitudes were used as detailed below. 
We maximised the detection efficiency by
performing systematic and comprehensive testing of the SExtractor settings 
with our data. Numbers of real and potentially
spurious objects were determined by using both deeper data, i.e. checking
the deeper optical images when performing NIR photometry, and negatives of the 
final images in case of optical frames.

We calculated numerous aperture magnitudes, as well as the AUTO and BEST 
magnitudes, which dynamically adjust the size of the aperture depending on the
shape and size of the object. By using curves-of-growth with bright 
non-saturated stars, we determined aperture corrections for slightly 
different seeing conditions and fixed size apertures.  
When giving colours 
we typically refer to magnitudes calculated inside 2.6\arcsec\ diameter 
apertures, approximately 2.5 times 
typical seeing, corrected to the $K$-band seeing value. 
However, total 
magnitudes, adopting the BEST magnitude which we find to be most robust 
in our uncrowded fields with
both extended galaxies brighter than $R$=23 mag and compact objects throughout
our magnitude range, are used when a single-band brightness is given for an 
object.  We checked these to be totally consistent with the total magnitudes 
derived from curves of growth used for standard stars.  
An exception is made with bright galaxies clearly larger than 
$\sim5$\arcsec\ for which these total magnitudes are used throughout, also
in the SEDs.
In summary, we estimate our photometry to be accurate
to approximately 0.03 mags in the optical and 0.05 in the NIR data.
Finally, we checked that our photometry was consistent within few percent 
with 2MASS galaxies found in our areas.

Zero-point calibrations were obtained through observations of standard 
stars throughout the nights, which were photometric. 
The absolute calibration to Vega-based
system is estimated to be accurate to 0.02 and 0.05 mag in the optical and 
NIR, respectively.  
The extinction coefficents were fixed to average values of the nights, 
to $0.08$, $0.05$, and $0.10$ in $R$, $I$, and $K$-bands, respectively.  
In addition, a colour-term of $0.1\times (R-I)$ was used to derive the
final optical magnitudes.

Since detection of optical objects was done with co-added $R+I$ frames,
the $R$ and $I$ band source lists have the same number of objects. These were 
cross-correlated with the $K$-band source list to produce source lists of
3-band detections, optical detections only (since $RI$ data is deeper than 
NIR), and NIR-only (including very red objects).

\subsection{SDSS data}

We also extracted {\em ugriz} photometry from the SDSS DR7 \citep{sdss}
archives for all our fields to complement our own dataset.  
SDSS data are shallower than our own optical $RI$ data, but
they extend the wavelength range of the brighter galaxies and thus make
SED fitting significantly more reliable.
After testing with different magnitudes we found that we acquire the best 
results, the smoothest combined SEDs, 
when adopting total magnitudes, as described above,
for our own photometry, and Petrosian magnitudes for the SDSS galaxies
brighter than $r<20$ mag, and SDSS model magnitudes for those fainter than this 
limit\footnote{see http://www.sdss.org/DR7/algorithms/photometry.html}.

\section{ISOPHOT sources in NGP fields}
\label{nature}

\subsection{Matching of FIR and optical/NIR sources}
\label{match}

As discussed in Section~\ref{intro}, identification of FIR sources is
difficult because of confusion and the large beam sizes.  Typical approaches
include using supplementary MIR or radio surveys in helping to decide 
which of the
many galaxies within the beam is the correct counterpart, and then running
likelihood ratio methods taking into account the distance of counterpart
candidates from the FIR source as well as the surface densities of all 
relevant populations.  Often the {\em one} MIR and/or radio galaxy giving 
the largest likelihood  is chosen as the
counterpart, unless {\em no} candidates are found above a given probability
threshold \citep[see e.g.][]{mann02,oyabu05,taylor05}. Further
analysis, SED fitting etc., is based on this identification.  
While the advantage obviously is to have a well-defined and quantitative
method, the downsides include possible biases depending on the ``indicator
source'' and especially the difficulty of assessing the
possibility of {\em multiple galaxies contributing to the FIR source}.
In this paper we attempt a much more straight-forward philosophy, 
though admittedly one less robustly defined.

We first searched for all sources with $R$, $I$, and $K$-band detections 
within 60\arcsec\ of
the centre of the ISOPHOT detection.  Making use of the SExtractor
star vs.\ galaxy neural network classifier, we excluded clear stellar 
objects by selecting only those with a mean CLASS parameter less than $0.75$ 
in all the three bands. Typically there remain 20 to 
50 extended objects per ISOPHOT error circle.  The optical/NIR sources were 
then matched also with the SDSS catalogs by searching for common objects 
within 1\arcsec\ radius.

\subsubsection{SED fitting and other analysis of candidates}

Next we fitted template SEDs using the HYPERZ software \citep{bolzonella}
to all sources found in the ISOPHOT error circles. 
The whole range of wavelengths from UV to FIR was fitted, 
our optical and NIR datapoints 
along with the five SDSS bands and three ISOPHOT points. 
The fainter targets not
having SDSS photometry were fitted with $RIK$+ISO only. 
The model templates were taken from, or calculated using, 
the GRASIL library/code \citep{silva98} -- in particular we used 
a range of Sc-type SEDs of different age, different age ellipticals, 
fully evolved Sa and Sb galaxies, as well as model fits to the SEDs of
NGC 6946, NGC 6090, M82, Arp 220, and HR10, which  respectively represent 
star-forming dwarfs, interacting/starbursting galaxies, warmer starbursts, 
(ultra) luminous IR galaxies, and extremely obscured objects.

When fitting the templates within
HYPERZ, we turned off reddening since the GRASIL 
SED models already have dust and gas properties with 
consistent extinction and re-emission mechanisms and
effects built in throughout the wide spectral range. 
We assumed a $\sim0.15$ mag uncertainty in the relative matching 
of FIR and ground-based parts of the SED due to uncertain aperture
effects.

The source catalogs in the target fields, now associated with 
photometric redshifts, best-fit SED templates, and especially the related
chi-square values and probabilities attached to the HYPERZ fits, form the basis 
of the selection of counterparts.  Note especially that we do {\em not} use
any information of the distance to the FIR detection centre in the analysis,
but treat all sources in the error circle equally. \citet{rodighiero05} 
cite a typical blended FIR source distance of 55\arcsec\ in their 90 $\umu$m 
data, and such cases would be severely misinterpreted by e.g.\ just picking
the closest candidate to the FIR detection centre \citep[see also][]{sajina06}.
On the other hand, we note that by fitting known templates to objects 
we cannot get away from the bias against totally unknown classes of sources, 
which plagues all FIR identification studies.

We also searched the SDSS spectroscopic archive for any available spectra
for the bright galaxies in the target fields.  
Within the ISOPHOT error circles
12 redshifts were found, and for 
these objects the redshift was fixed in the HYPERZ fit.

To have a quantitative measure of the morphology of the brighter galaxies in 
the field, we used the GALFIT program \citep{peng02} 
to model the surface brightness distributions of the $RI$ combined images.

\subsubsection{Breakdown of the candidates}

Based on the types of optical-NIR objects within the 1 arcmin search 
radius characterized by their magnitudes, SEDs, model fits and redshifts
(photometric or spectroscopic), as well as morphologies,
we separated the 22 FIR objects in the present sample into
two groups:  In the first instance, we find 5 fields which have 
a single bright $K\sim14$ mag galaxy in the ISOPHOT detection 
circle.  All these galaxies are more than 
two magnitudes brighter than any other object in the area and they are 
all well fit with a template SED all the way from UV and optical 
to FIR. They also
have SDSS spectroscopy available which is consistent 
with their nature as FIR sources.  They are all brighter than $K$=14.8 mag and
we note that using average bright $K$-band galaxy counts
\citep[e.g.][]{petri00}, there is less 
than 10 per cent probability to find a $K<14.8$ mag galaxy inside a 60\arcsec\ 
radius circle. 
We refer to these five {\em unambiguous ISOPHOT counterparts} 
as the ``bright sample''.

The second group of ISOPHOT targets, 17 of them, is called the 
``faint sample'', referring to the brightness of the counterpart 
galaxies, not that of the FIR flux. There are essential differences 
amongst these sources, however.  More than half of the fields do actually
have fairly bright $K\sim13-16$ mag galaxies in them, but there are 
usually two to four of them, suggesting possible blending. 
The rest of the fields, on the other hand, look initially empty of any
obvious counterpart candidates.
When investigating the faint sample individually below, we categorize them
according to how and why the cases are ambiguous.

\subsubsection{Further modelling of SEDs}

To have an independent characterization of the brighter counterparts
with redshifts, or secure estimates thereof, we modelled them 
with the radiative transfer models of  \citet{andreas00} and 
\citet[][]{andreas03} (hereafter ERR03). The models incorporate
the stellar population synthesis model of \citet{bruzual} and the dust
model of \citet{sieben92} which treats small transiently
heated grains and PAH molecules as well as large classical grains. The
models have been shown to be in good agreement with the spectral energy
distributions of starburst and cirrus-dominated galaxies 
\citep{andreas00,andreas03,farrah03,taylor05}.

\subsection{Unambiguous bright galaxies}
\label{brightgals}

\begin{figure*}
   \centering
   \includegraphics{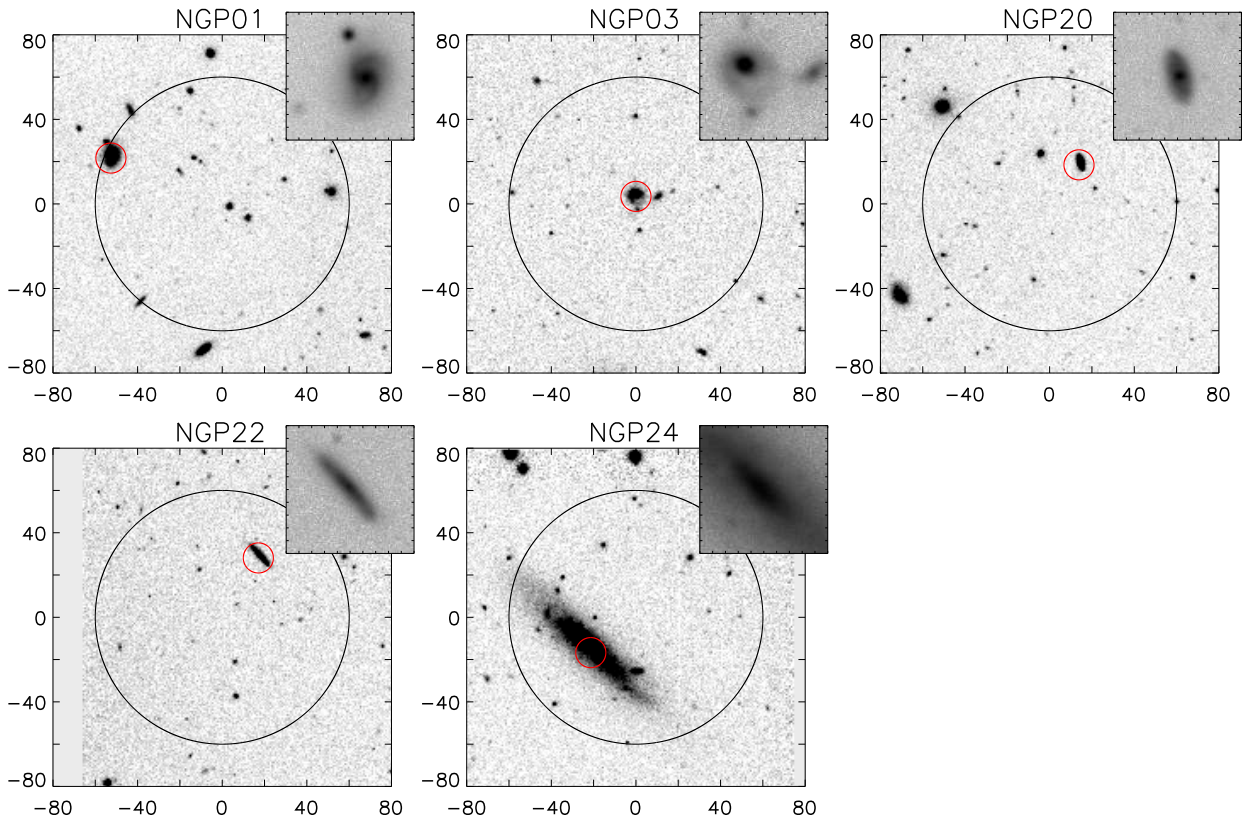}
   \caption{Optical $R+I$ combined images of the bright sample of the 
ISOPHOT NGP source fields.  The large circle of one arcmin radius
shows the location of the $ISO$ detected source.  The small red circle
identifies the likely optical counterpart discussed in the text and
zoomed-up in the inset -- its best-fit SED template is 
shown in Fig.~\ref{SEDs-bright}. North is up and East left.}
              \label{FCs-bright}%
    \end{figure*}

\begin{figure*}
   \centering
   \includegraphics{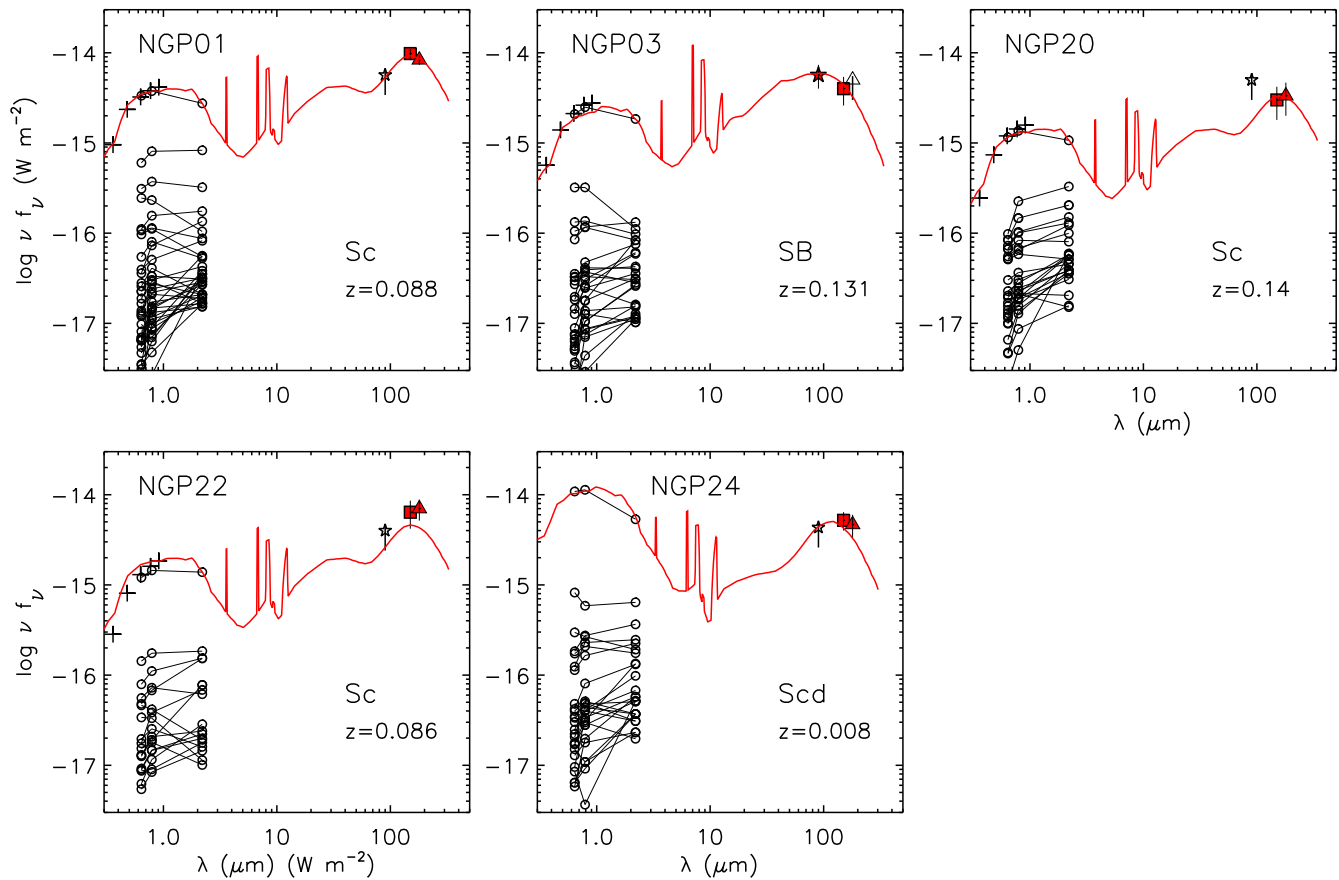}
   \caption{SEDs of bright individual galaxies in CIRB ISOPHOT fields.  
Circles represent our $RIK$
photometry within 60\arcsec of the FIR detection, and the crosses show SDSS 
$ugriz$ data of the bright counterpart. 
The stars, squares, and 
triangles are the 90, 150, and 180 $\mu$m ISOPHOT data, respectively - filled
symbols mark detections, whereas unfilled symbols show an upper limit.
The best fit GRASIL SED is overplotted as the red solid line, and the type
with the redshift if indicated.  
}
              \label{SEDs-bright}%
    \end{figure*}

All the FIR sources of the bright sample defined above have
FIR source quality flag values of $q\geq3$ \citep[see][]{juvela00} in at 
least one detected band. 
The sources are listed in Table~\ref{table-bright}
along with some characteristics based on GRASIL model fits, whereas 
Table~\ref{table-bright2} lists the ERR03 model fits.
Based on our optical imaging, all these sources are disk 
galaxies.  Images from the combined $R+I$ ALFOSC imaging are shown in 
Fig.~\ref{FCs-bright}.  SEDs of all sources within 60\arcsec\ radius 
from the ISOPHOT source are shown in Fig.~\ref{SEDs-bright} and the 
best-fitting SED to the bright candidate is also overplotted.  
The detailed description of each of the five fields, with discussion 
leading to the identification of the source, its SED and 
morphological fit, is presented in the Appendix.  

As can be seen from the Tables~\ref{table-bright} and~\ref{table-bright2} 
the likely counterparts are generally nearby $z<0.2$ (strongly) star-forming 
late type spiral galaxies. The general characteristics
of the population is discussed below in Section~\ref{brightdiscuss}.

%

   \begin{table*}
      \caption[]{Bright unambiguous ISOPHOT counterparts are listed 
with their SDSS redshifts, the best fitting 
UV-to-FIR GRASIL SED template from HYPERZ, the absolute $K$ magnitude,
and effective radius in kpc from GALFIT. 
$L_{IR}$ is the log of IR luminosity, in solar units, 
integrated over $8-1000 \ \umu$m of the best-fit GRASIL model and SFR is 
calculated from this using the FIR/SFR relation of \citet{kennicutt}.}
         \label{table-bright}
     $$ 
         \begin{tabular}{llllrrrrl}
            \hline
            \noalign{\smallskip}
            FIR source   &  counterpart & $z$ & SED & M$_{K}$ 
                         &  $L_{IR}$ &  $r_s$ & SFR &  comments \\
            \noalign{\smallskip}
            \hline
            \noalign{\smallskip}
	    NGP01 & 2MASSXJ13412738+4042166 & 0.088 & Sc$^a$ & -24.22 
                         & 10.94 & 2.9 & 15 & disturbed \\
	    NGP03 & 2MASSXJ13422216+4022017 & 0.131 & SB$^b$ & -24.72 
                         & 11.17 & 6.3 & 26 &  interacting \\
	    NGP20 & J134934.9+390730.1$^c$  & 0.140 & Sc$^a$ & -24.28
                         & 10.92  & 2.9 & 14 & undisturbed \\
	    NGP22 & SDSSJ135054.71+385847.2 & 0.086 & Sc$^a$ & -23.42 
                         & 10.62 & 3.6 & 7 & disturbed \\
	    NGP24 & UGC 08793 &  0.0081 &  Scd$^d$ & -19.63 
                         & 8.49 & 2.2 & 0.05 & warped, many bright HII regions \\
            \noalign{\smallskip}
            \hline
         \end{tabular}
     $$ 
\begin{list}{}{}
\item[$^{\mathrm{a}}$] nominally 15 Gyr Sc-model
\item[$^{\mathrm{b}}$] NGC 6090 model
\item[$^{\mathrm{c}}$] uncatalogued - see text 
\item[$^{\mathrm{d}}$] NGC 6946 model
\end{list}
   \end{table*}
%

%
%

   \begin{table*}
      \caption[]{The bright ISOPHOT counterparts fitted with
models of \citet{andreas00,andreas03}. 
The logs of the cirrus and starburst IR luminosity 
components, $L_{C}$ and $L_{SB}$, respectively, in solar units, 
do not necessarily
equal the total listed in Table~\ref{table-bright} based on GRASIL models, 
though the results are clearly consistent in each case.
All starburst models assume an age of 26 Myr and
an initial A$_V$ of the molecular clouds of 50. We also list the 
resulting peak star-formation rate, and the log of dust mass in solar units 
and the temperature in K.  }
         \label{table-bright2}
     $$ 
         \begin{tabular}{llrrrrrr}
            \hline
            \noalign{\smallskip}
            FIR source   & $z$ & $L_{SB}$ &  $L_{C}$ & A$_V$ 
            & SFR & T$_{dust}^{a}$ & log M$_{dust}^{b}$   \\
            \noalign{\smallskip}
            \hline
            \noalign{\smallskip}
	    NGP01 & 0.088 & 10.66 & 11.16 & 1.3 & 20 & 17 & 8.3  \\  
	    NGP03 & 0.131 & 10.68 & 11.15 & 1.1 & 22 & 25 & 7.5  \\  
	    NGP20 & 0.140 & 10.62 & 11.07 & 1.3 & 20 & 17 & 8.3  \\  
	    NGP22 & 0.086 & 10.55 & 10.95 & 1.3 & 17 & 17 & 8.2  \\  
	    NGP24 & 0.0081 & --   &  9.16 & 0.3 & -- & 17 & 5.9  \\  
            \noalign{\smallskip}
            \hline
         \end{tabular}
     $$ 
\begin{list}{}{}
\item[$^{\mathrm{a}}$] The temperature of the dominant 0.24$\mu$m graphite grain 
                      species in the best-fitting model in the FIR is listed.
\item[$^{\mathrm{b}}$] Dust mass is estimated as in \citet[][]{taylor05} but
the method has been generalised to take into account the distribution of dust 
temperature predicted by the model. 
\end{list}
   \end{table*}
%

\subsection{Confused fields}

\begin{figure*}
   \centering
   \includegraphics{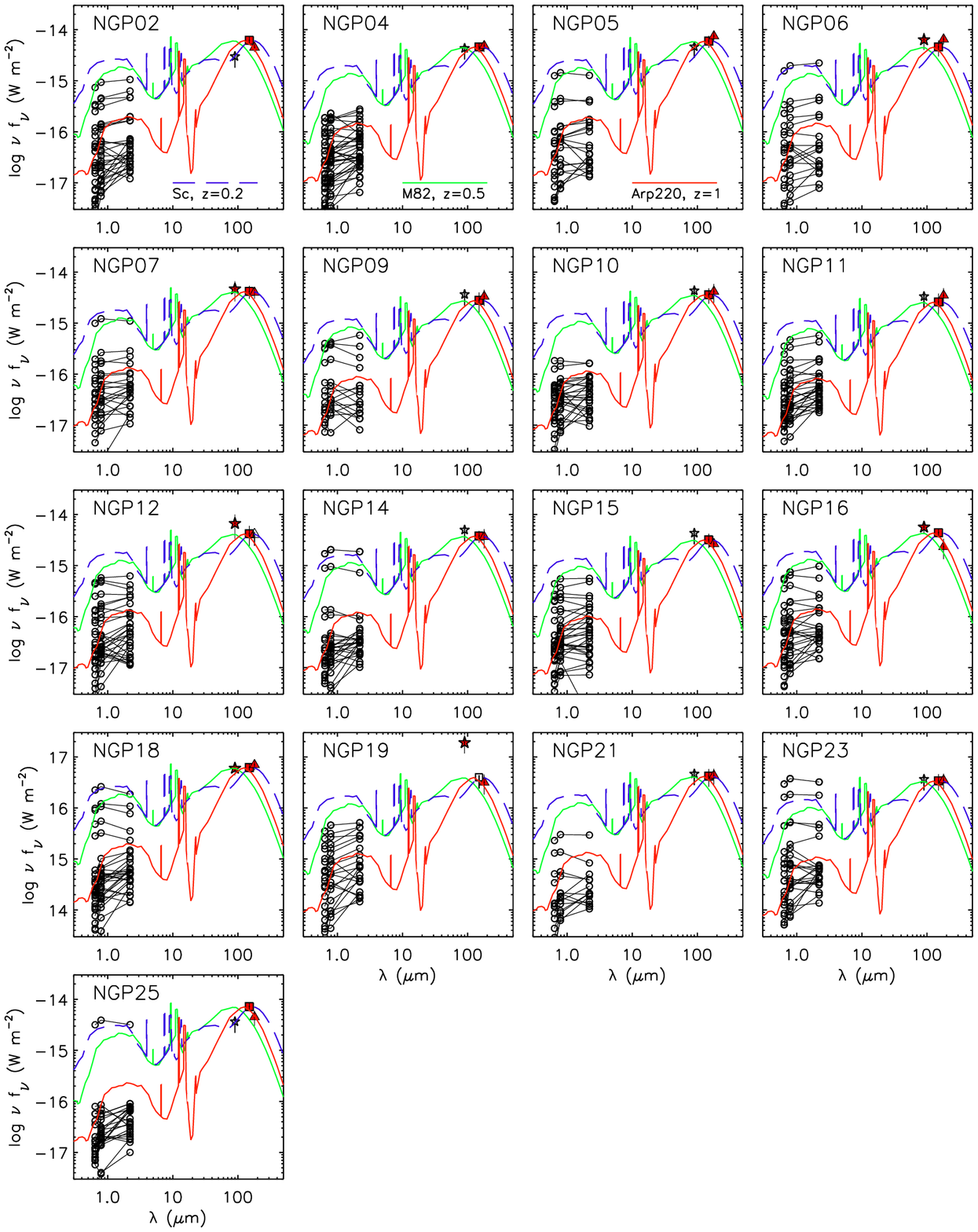}
   \caption{SEDs of galaxies in ISOPHOT fields containing more than one 
good candidate for the FIR counterpart.  For clarity, we overplot only
Sc, M82 and Arp 220 SEDs normalized to the 150 $\mu$m flux at fixed 
redshifts. See text for discussion on individual cases.
}
              \label{SEDs-medium}
    \end{figure*}

There are 17 cases 
in the present FIR sample where we cannot find a single unambigious
counterpart, though in many cases there are fairly bright, $K=13-16$ mag
galaxies in the vicinity. 
In fact, in more than half of the fields 
there are two to five $K<16$ mag galaxies within the ISOPHOT detection error 
circle, while statistically only 0.4 $K<16$ galaxies are expected.
It is likely that the
majority of these cases are indeed confused FIR sources 
where the IR flux originates from several individual sources.  
Images of fields based on ALFOSC 
data are shown in Fig.~\ref{FCs-medfaint1}. We discuss the cases 
individually in the Appendix, while 
a summary is presented in Table~\ref{mediumtab}, and the general
physical characteristics are discussed in Section~\ref{confdisc}. 
Figure~\ref{SEDs-medium} shows the $RIK$+FIR SEDs of the target fields;
note that the overplotted SED templates are indicative of typical templates
and typical redshifts, and {\em not} the best 
fits for the individual cases.

   \begin{figure*}
   \centering
   \includegraphics{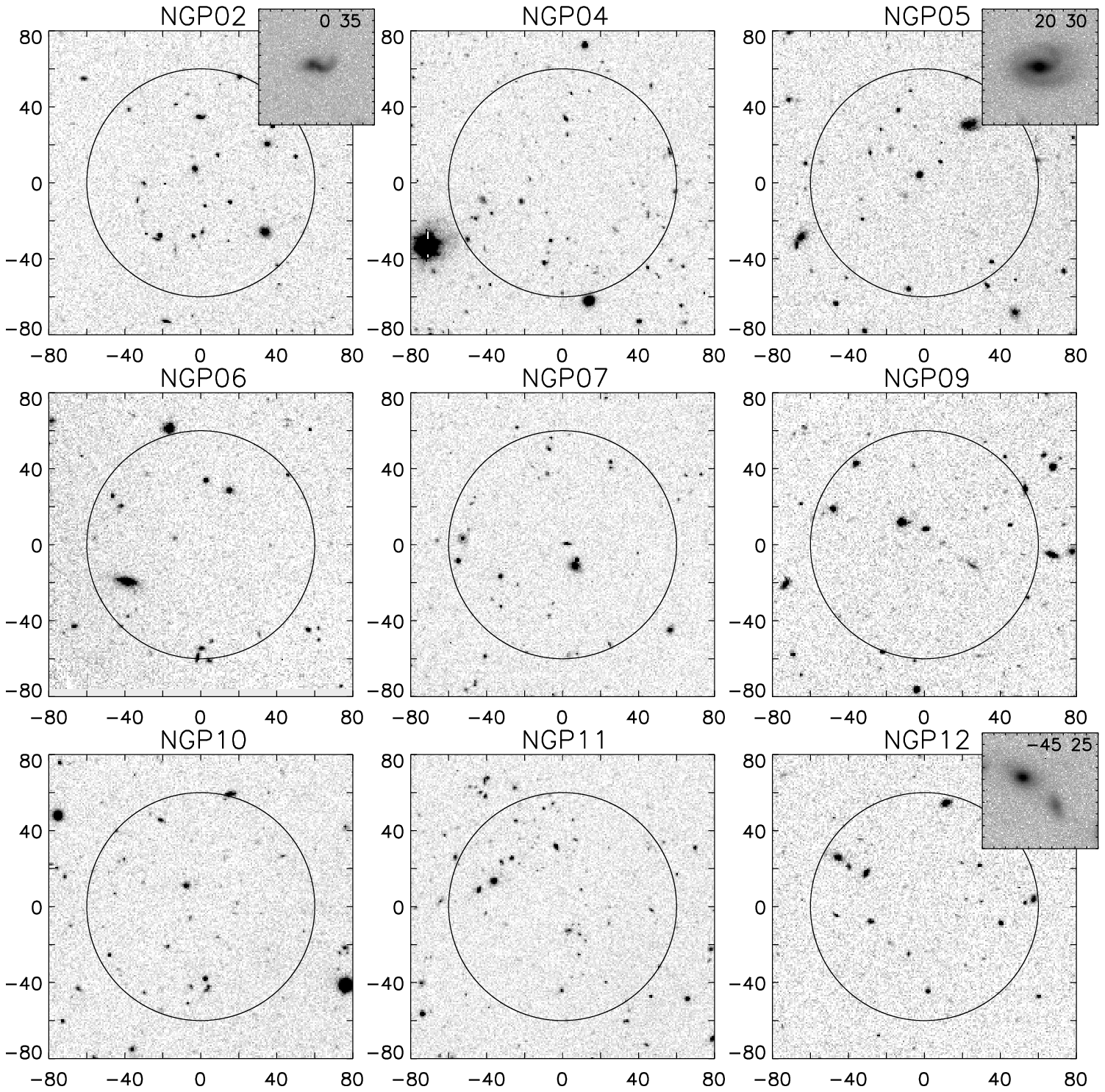}
   \caption{Optical $R+I$ combined images of the faint sample of the 
ISOPHOT NGP source fields, where multiple and/or faint and distant counterparts
are identified.  The large circle of one arcmin radius
shows the location of the ISO detected source.  
If one of the counterpart candidates is a disturbed or 
interacting galaxy at $z<0.3$, it is shown in the inset, and its {\em x,y}
position in the main image is given.}
              \label{FCs-medfaint1}%
    \end{figure*}
   \begin{figure*}
   \centering
   \includegraphics{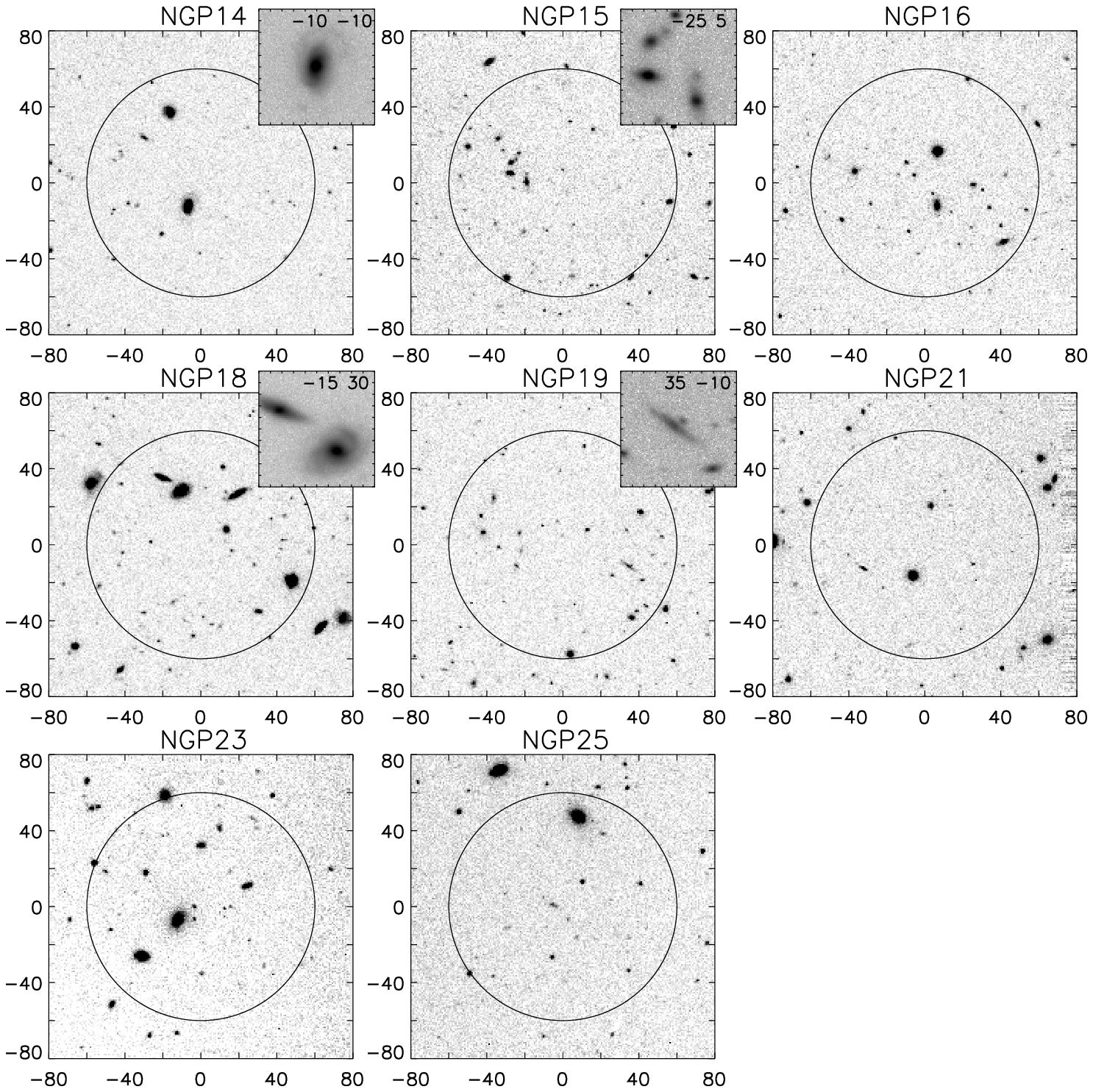}
   \contcaption{}
    \end{figure*}

   \begin{table*}
      \caption[]{The 'confused sample' of the ISOPHOT counterparts.
                 We classify the sample according to the likeliest scenario
                 for confusion: type-1 -- sum of several bright
                 galaxies; type-2 -- sum of bright and faint galaxies;
                 type-3 -- sum of several faint galaxies;
                 type-4 -- one of several possible counterparts.
                 }
         \label{mediumtab}
     $$ 
         \begin{tabular}{llrll}
            \hline
            \noalign{\smallskip}
            FIR source   &  $N(K<16)$ & type  & summary  & other comments \\
            \noalign{\smallskip}
            \hline
            \noalign{\smallskip}
	    NGP02 & 4  &  1 & 2+ $z\sim0.3$ spirals   &  \\
	    NGP04 & 0  &  3 & faint (U)LIRGs &  $z=0.7$ cluster in area\\
	    NGP05 & 3  &  1 & 2+ $z\sim0.25$ spirals &  \\
	    NGP06 & 3  &  1,2& $z=0.16$ SB + faint (U)LIRGs &  \\
	    NGP07 & 1  &  1 & 2+ $z\sim0.3$ spirals  & merged ISOPHOT detection  \\
	    NGP09 & 1  &  3 & faint (U)LIRGs & low quality ISOPHOT detection \\
	    NGP10 & 0  &  3 & faint (U)LIRGs &  \\
	    NGP11 & 2  &  1,2 & 2 $z\sim0.3$ spirals + faint (U)LIRG &  \\
	    NGP12 & 2  &  1,4& 2+ $z<0.2$ spirals & merged ISOPHOT detection \\
	    NGP14 & 2  &  1 & 2 $z=0.16$ and $z\sim0.1$ SB/spirals &  \\
	    NGP15 & 2  &  3 & 2+ $z\sim0.3$ SBs and/or faint (U)LIRGs & low quality ISOPHOT \\
	    NGP16 & 2  &  1,2,4& 2 $z\sim0.3$ SBs and/or faint (U)LIRGs & merged ISOPHOT detection\\
	    NGP18 & 4  &  1 & 2+ $z=0.13$ spirals  &  \\
	    NGP19 & 3  &  1,2,4& 2+ $z\sim0.3$ SBs and/or faint (U)LIRGs &  \\
	    NGP21 & 0  &  2,3& faint (U)LIRGs  & low quality ISOPHOT detection \\
	    NGP23 & 4  &  1,4& 1-2 $z=0.2$ spirals/SBs &  \\
	    NGP25 & 1  &  1 & 2+ $z<0.2$ spirals & bright companion outside area \\ 
            \noalign{\smallskip}
            \hline
         \end{tabular}
     $$ 
   \end{table*}

\section{Discussion}

\subsection{The nature of unambiguous counterparts}
\label{brightdiscuss}

We found in Section~\ref{brightgals} that five
of our 22 FIR sources, i.e.\ 23 per cent, are unambiguously identified 
with low-redshift $z<0.2$ star-forming spiral galaxies.
Apart from NGP24 all are just about at or above the LIRG luminosity
criterion of $>10^{11} \ {\rm L}_{\sun}$.  
Apart from NGP20 all have evidence of disturbed morphology and
NGP03 
is the clearest case of an interacting system. 
Assuming typical $K$-band mass-to-light ratios 
of 
star-forming spiral galaxies \citep{bell,gildepaz} the estimated
stellar masses range from 
4 to $15\times 10^{10} \ M_{\sun}$, i.e.\ close to a mature 
$m_{\star}$ galaxy \citep{cole01}.
The exception in this class is NGP24, a very nearby, 34 Mpc, 
star-forming dwarf galaxy 
with an absolute brightness of a mere $M_K=-19.6$. 

Should this set of galaxies, ignoring NGP24 from now on,  
be characterized as normal quiescent spirals or starbursts?  
The SFRs derived by both GRASIL and ERR03 fits are in the range of 
$\sim 7-26 \ M_{\sun} \ yr^{-1}$, i.e.\ higher than ordinary local spirals by 
factors of few, but still lower than typical strong nuclear starbursts.  
The ERR03 models, 
Table~\ref{table-bright2}, show them to be cirrus dominated, but with 
significant contributions from warmer dust in 
star-bursts.  Determination of the specific star-formation rate (SSFR) 
also place our galaxies somewhere in between quiescent galaxies and starbursts:
SSFRs are $\log (SFR/M_{stellar}) \sim -9.8$, 
similar to local $H\alpha$ selected
star-forming galaxies \citep[see e.g.][]{gildepaz,sajina06}.
The ISOPHOT sample 
SSFR values are clearly higher than those of local galaxies in general, 
even the `blue' ones: most of the stellar mass in the local Universe peaks
at $\log (SFR/M_{stellar}) \sim -11.6$ while the blue cloud is at
around $\log (SFR/M_{stellar}) \sim -10.2$ \citep[e.g.][]{kauffmann04}.
Note some ambiguity in the term ``starburst'': 
some studies find ``starbursts'' 
while others find galaxies with ``slightly elevated star-formation'' compared
to normal local spirals, while actually meaning similar kind of galaxies.

The peaks of the rest-frame SEDs are in the range 
100 to 150 $\umu$m.
According to the ERR03 modelling the galaxies have cold dust temperatures 
in the range 15 to 20 K -- these values refer to the dominant dust 
population, the fits are {\em not} fitted with a single dust species and 
temperature.  Dust masses are quite uniformly $\sim 10^8 \ M_{\sun}$.
IRAS $\log(f_{\nu}(100\umu m) / f_{\nu}(60\umu m))$ and 
Spitzer $\log(f_{\nu}(160\umu m) / f_{\nu}(70\umu m))$ colours of the 
unambiguous sources, as calculated from the best-fit models are in the range
-0.3 to -0.5, and -0.3 to -0.7, respectively.  These values are
typical to the infrared properties of the more quiescent FIR-selected galaxies, 
rather than active ones, in the nearby Universe 
\citep[e.g.][]{dale01,sanders03,dale05}. Hence, the bright ISOPHOT galaxies 
in this field appear to be 
gas rich galaxies with clearly elevated star-formation at about the LIRG-class
limit, but without evidence in our data for strong (nuclear) star-bursts.
From optical data alone, e.g.\ with their quite uniform 
$g-r \approx 0.75$, our sample is indistinguishable from other local 
relatively high mass ``blue-cloud'' galaxies. As they are close to the 
red-sequence, it could be speculated that they are moving towards it, the 
effects of which can be seen in their IR and morphological properties. 

The bright FIR selected sources in this survey and others tend to be 
(nearby) cold galaxies, whereas the local 
ULIRGs typically have warmer SEDs.  However, at higher redshifts where LIRGs
and ULIRGs start to dominate the energy budget of the Universe, Spitzer
FIR selected galaxies in fact appear to have colder 
SEDs \citep[e.g.]{zheng07,symeonidis09}, 
more similar to the majority of local FIR-bright galaxies 
rather than local ULIRGs.  This makes detailed studies
of the physical conditions in local FIR selected samples of IR-bright 
{\em non-}ULIRGs very motivating.

\subsection{The nature of confused counterparts}
\label{confdisc}

Among the rest of our FIR targets, the faint sample, 
there are several cases where there either are optional counterparts, or
strong reasons to suspect a blend of two or more bright galaxies.  The (likely)
properties of this set can be summarized as follows:

A blend of two or more spiral galaxies at redshifts $z<0.3$
is responsible for the ISOPHOT detection in 9 cases (NPG02, NGP05, NPG07, NGP11,
NGP12, NGP14, NGP18, NGP23, NGP25). 
The photometric redshifts and GRASIL SED types
of the individual counterparts are essentially identical to the unambiguous 
cases in Sections~\ref{brightgals} and~\ref{brightdiscuss}, 
though the optical
colours show a slightly larger range than the unambiguous bright sample. 
Since we are not able to distribute
the FIR flux to blended counterparts with spectroscopic redshifts, 
we will not attempt to model their physical characteristics in detail.  
However, we did run the ERR03 models on several cases where there e.g.\ were 
two spiral galaxies with reasonably secure photometric redshifts:  the 
cirrus and starburst luminosities and dust characteristics and
star-formation rates again come out to be
very similar to the bright unambiguous set of FIR sources.
They are often at slightly higher redshifts, but due to dividing the FIR
fluxes to two or more sources, the $L_{IR}$ and SFR values end up very 
similar. Thus, in total {\em 14/22 or
64 per cent of the FIR sources in the ISOPHOT EBL project NGP fields
are low-redshift $z<0.3$ moderately
star-forming galaxies}. The discussion presented in 
Section~\ref{brightdiscuss}
above applies directly to all these sources, the majority 
detected in our survey.  

Four other blended FIR sources (NGP06, NGP15, NGP16, NPG19) also 
include a contribution, possibly a major contribution, from a bright spiral or 
starburst.  In these fields, however, we
have also identified fainter red galaxies which might well be higher redshift 
(U)LIRGs also contributing the FIR flux.

\subsection{Higher redshift counterparts}

There are four fields (NGP04, NGP09, NGP10, NPG21), where the only likely 
counterpart appears to be a higher redshift, $z>0.4$, (U)LIRG, or a sum of 
them.  Higher redshift counterparts thus account for a minimum of 18 per cent 
of the FIR sources.  Including ones where a higher-$z$ source is possibly
blended with a lower-$z$ counterpart, up to a maximum of 36 per cent of our
NGP targets include an IR-galaxy at $z>0.4$.

\subsection{Comparing to previous surveys and models}

Overall we found 23 per cent of the NGP fields to have a uniquely identified 
target, while another 41 per cent are identified as blends of bright galaxies.
In the various FIRBACK and Lockman Hole surveys, typically 
$\sim50$ per cent of sources are identified \citep[][]{dennefeld05,taylor05,
oyabu05,rodighiero05}, and the rest remain unidentified. 
The fractions are
similar to ours even though the methods of identification have been quite
different: most of the above works considered only a bright sub-set, 
typically half, of their total source catalogs, and they all employed 
mid-IR and radio data in the identification which we have not done.  

If the general fractions of identified and unidentified sources are
similar to other works, what about the characteristics of the counterparts?
We identified at least 64 per cent of all targets as fairly normal cool or
cold IR luminous spiral galaxies at $z<0.3$, regardless of confusion issues.  
If the cases of likely bright and faint galaxy 
blends are included, the fraction rises to 82 per cent.
Similarly, the fraction of contributions from higher redshift (U)LIRG type 
galaxies is in between 18 and 36 per cent.  It is important to note that
these cases are not different in their FIR properties from the rest of the 
NGP ISOPHOT sources.
Again, the results are very consistent with other surveys. 
In general, previous ISOPHOT FIR 
follow-up surveys have found a bi-modal distribution of a large, approximately
three-quarters majority of 
low-redshift quiescent galaxies and a minority higher redshift 
ULIRGs at 
$z=0.4-0.9$ \citep[e.g.][]{sajina03,patris03,taylor05,rodighiero05,
dennefeld05,sajina06}.  
In the studies probing somewhat fainter flux levels of 100 mJy 
\citep[e.g.][]{oyabu05} than ours and FIRBACK's $\sim$150 mJy, the fraction
of sources at redshift higher than 0.3 starts to rise slightly.  
The dust temperatures we find are similar to \citet[e.g.][]{sajina06} 
and \citet{patris03} while
\citet{taylor05} derive slightly higher temperatures in the range 20--40 K.
Early results from the FIR surveys with {\em AKARI} also suggest large 
contributions from local $z<0.1$ galaxies \citep{matsuura07,malek09}.

Spitzer MIPS results
also result in a broadly equivalent picture: the deeper MIPS sources at
70 $\umu$m (down to 4 mJy) appear to be starbursts at $z$=0.1--1.2 with a 
mean and median at $z\sim0.5$ \citep[e.g.][]{symeon08}, while the 
160 $\umu$m sources (to 100 mJy) are mostly 
``starburst galaxies'' at $z\sim0.2$ though
generally not of the active, warm SED type of M82 and ARP 220, 
but with cooler SEDs \citep[e.g.][]{frayer06b,wen07}.
  
We do identify a difference to previous follow-up surveys, however.  
Four out of five of our bright unambiguous counterparts, as well as more 
than half of the identifiable blended counterpart options are clearly 
disturbed or interacting galaxies.  
In contrast, \citet[e.g.][]{sajina06} found that their counterparts
to ELAIS N1 survey do {\em not} show evidence of interactions, and also
\citet{oyabu05} cite a small minority as interacting sources in the Lockman 
Hole survey.  We are not
sure what the reason for this difference is, unless it is just that our
optical data is somewhat deeper than theirs making morphological 
classification easier.

The majority of the galaxy counterparts in ``deep'' FIR surveys 
discussed in this paper are in fact very similar to the large sample of 
much brighter galaxies at $>2$Jy detected as part of the ISOPHOT 170 $\umu$m 
Serendipity Survey \citep{stickel07}.

\subsection{Models}

Previous FIR surveys
have been described to a reasonable accuracy by models of  
e.g.\ \citet[][]{lagache04} and recently by \citet[][]{rowan09}. The models
have been specifically constructed to fit the source counts at various
wavelengths, and also the level and spectral shape of CIRB.  
They are broadly consistent with the 
bimodal distribution of FIR sources described above. 
For example the 160 $\umu$m Spitzer counts
are dominated by ``cirrus'' galaxies brighter than 80 mJy, and by M82-type
starbursts fainter than this limit
according to the \citet[][]{rowan09} models, and at 150 mJy, more or less at
our survey limit, quiescent galaxies should outnumber starbursts by a factor
of three, consistently with the nature of $ISO$ sources found in the NGP
fields.
However, a recent FIR/sub-mm follow-up of $BLAST$ at the longer
wavelengths of 250 and 500 $\umu$m \citep[][]{dye09} find that these same
bimodal models are a poor fit to the their data, as are also 
early AKARI counts of \citet{matsuura07}, 
highlighting the fact that
the exact nature of galaxies emitting their peak at 
100 to 200 $\umu$m, where the CIRB also peaks, remains of great 
importance.

\subsection{Confusion at FIR wavelenghts}
\label{confused}

In our NGP-field follow-up most of the FIR sources, 
17/22 or approximately 70 per cent, 
cannot be identified unambiguously with a single
optical counterpart, though there clearly are different classes of ambiguity.
Of these 17 ISOPHOT detections 9 are cases where the most likely 
counterpart of the FIR detection is a sum of two or more 
fairly bright ($K<16$ mag) and nearby ($z<0.3$) star-forming galaxies, and in
another four cases one such bright nearby spiral appears to be 
a component in the confused FIR source.  Further four fields do not have any 
nearby galaxies in them, leaving the explanation for the FIR flux to either 
be a single higher redshift ULIRG type source, or a sum of several of these -- 
there is no way to differentiate between these cases with the present data. 
 
Thus, based on the present data, about 60 per cent of 
whole sample are definitely
blended objects, with another 20 per cent possibly so.
Similarly to our survey, \citet[][]{dennefeld05} find 50 per cent of their
identified sources multiple, and \citet[][]{sajina06} note that up to 
50 per cent of the FIRBACK sources are not ideally matched 
with a single counterpart.

\subsubsection{Effect on source counts}

\begin{figure*}
   \centering
   \includegraphics{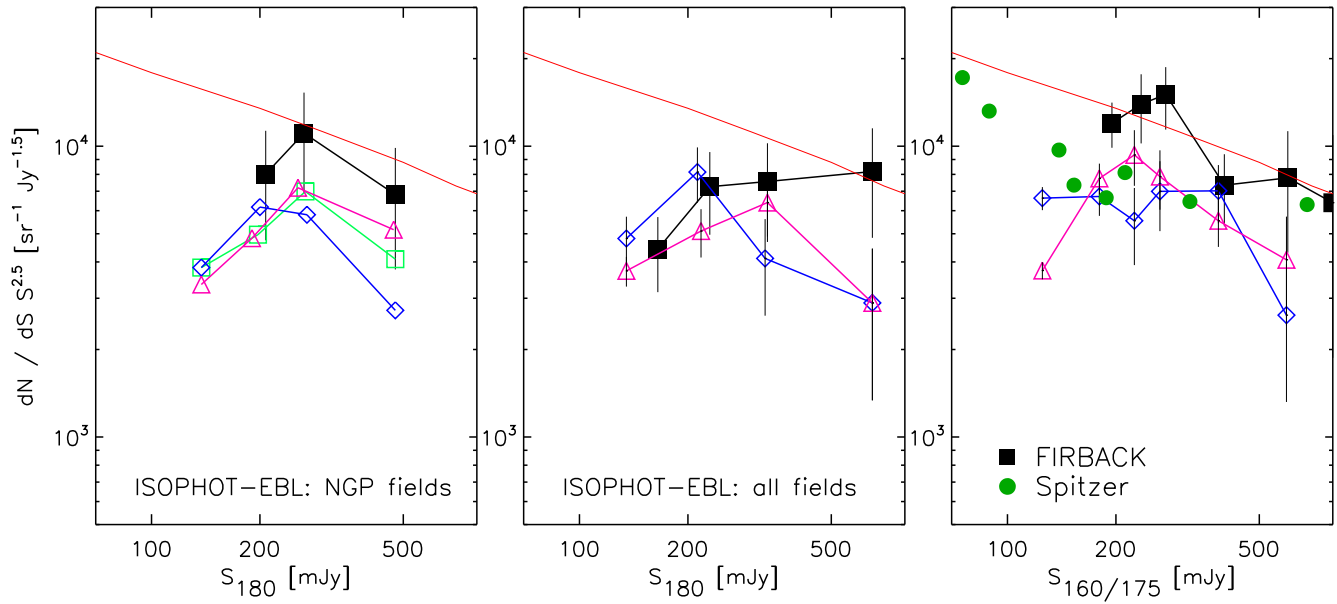}
   \caption{Uncorrected differential source counts in the 180 $\umu$m band, 
normalized with a Euclidean slope, 
are shown with solid symbols for the NGP fields at left, 
all ISOPHOT EBL Project areas in the
middle, and ELAIS N1 and N2 fields on the right.
Open symbols show effects of blended sources on the source counts; see 
text for details of confusion and blending effects.  
A model prediction from \citet{lagache04} is shown for
reference as the red curve.  The right panel also shows the Spitzer/MIPS
160~$\umu$m counts of \citet{frayer06b,frayer09}, which reach a factor
of 2-3 fainter than the ISOPHOT counts displayed, as green circles.}
              \label{counts-v1}%
    \end{figure*}

How does the blending and confusion affect interpretations of FIR 
populations? 
In addition to making identifications and SED fits of individual galaxies 
difficult and sometimes suspect, it also affects the FIR source counts 
themselves. As an example, Fig.~\ref{counts-v1} shows the raw 180~$\umu$m 
source counts from our ISOPHOT EBL project \citep[][]{juvela00} as well
as raw FIRBACK/ELAIS counts from \citet[][]{dole01}.
The solid symbols are original points, whereas the open symbols show the 
effect of blending.  In the left panel 
we plot the NGP counts, with the following
modifications: green open 
squares are a case where the 9 objects definitely consisting of
bright blends are divided into two discrete equal-flux sources; blue 
diamonds show a case where each detection in the
faint sample is divided into two discrete sources of equal 
flux; and magenta triangles depict a case where each detection in the faint 
sample is divided 
into two sources where one has 2/3 and the other 1/3 of the flux.

To see the effect with somewhat larger number statistics, the second panel 
shows the counts from the whole ISOPHOT EBL survey.  In this case we do not
have the information on possible blending of each individual source, so we make
a statistical estimate. The blue open
diamonds, equivalently to the left panel, 
show the average of 10000 realizations of the original counts,
where randomly selected 60 per cent of the FIR sources 
are replaced with {\em two} sources of half of the original flux each.
The error bars reflect the standard deviation of the simulated counts.  
The triangles show the case where the random 60 per cent are divided
into a 2/3 and a 1/3 flux source, as above.
We remind the reader that there is no difference in the flux distribution
of the blended and unique sources, justifying the selection of split
sources from the whole population.
Since FIRBACK/ELAIS surveys saw very similar fractions of non-unique 
counterparts 
\citep[e.g.][]{dennefeld05,sajina06}, we ran the same test on the FIRBACK
175 $\umu$m catalog, and the result is shown in the right panel of 
Fig.~\ref{counts-v1}, where the symbols and colours are equivalent to the
first two panels. In addition, we overplot the raw MIPS 160 $\umu$m counts 
of \citet{frayer06b,frayer09} with green circles. Only those points in the 
FIRBACK and MIPS counts  which have completeness of 85 per cent and above 
are plotted.

Though the faintest plotted raw count bins are affected by incompleteness, 
it still is clear from Fig.~\ref{counts-v1} that the reshuffling of flux 
typically takes some 30 to 60 per cent of the abundant 100 to 400 mJy 
sources and shifts them into flux bins in between 50 and 200 mJy. The 
resulting differential counts move down broadly by a factor of $\sim1.5$, 
and steepen the faint end in the presented examples where 
60 per cent of sources are affected . We ran the 
statistical blending correction with a variety of affected percentage of 
sources, and a range in the fractions and number of underlying sources 
corresponding to a FIR detection:  e.g.\ when 30 per cent of sources are 
affected in a way that 1/5 of a source flux is given to a new faint source, 
the necessary downward correction of counts is still approximately a factor 
of $1.2$.  Where splitting is more even the correction is larger, and rises
to factors of few if three or more underlying sources are concerned. Only if the
affected source population is below 20 per cent, are the required corrections
negligible compared to the errors of the observed counts.

Interestingly, the MIPS counts fall closer to the blending-corrected 
ISOPHOT counts and are not compatible with the uncorrected ones. 
The explanation is in the confusion levels. There
are two kinds of ``confusion'': one due to fluctuations of all the unresolved
sources below the detection limit and the other due to the effective beam
size not resolving properly sources too close to each other. One can talk
about confusion ``noise'' in the first case, and e.g.\ ``blending'' in the 
second case, and of the respective confusion limits as a ``photometric 
confusion limit'' and a ``source density criterion'' often given in terms
of beams per source in the latter case.  
Both effects are non-trivial as they obviously depend on the shape of the 
source count slope both above and below the confusion limits and other factors 
\citep[see e.g.][and references therein]{petri01,dole03,takeuchi04}.
For the cases here, the MIPS 160 $\umu$m beam 
of $\approx40$\arcsec\ is smaller than the respective C200 ISOPHOT beam of
$\approx90$\arcsec\ size, and assuming a Euclidean slope this means that the 
confusion ``noise'' is expected to be 3 times higher for ISOPHOT, and
even more for steeper counts. Indeed, for a set of source counts \citet{dole03} 
and \cite{jeong06} estimate the confusion ``noise'' limits to be in the region
of 50 and 160 mJy for MIPS 160 $\umu$m and ISOPHOT 180 $\umu$m, respectively.
These values correspond roughly to $\sim40$ or more beams per source, 
which is a limit one should use with steep counts \citep{petri01}. 
Hence, the ISOPHOT counts we are considering 
clearly reach right to the level of expected confusion, whereas only
the two deepest \citet{frayer09} points at $< 100$ 
mJy in the Fig.~\ref{counts-v1} approach the MIPS confusion limit. 

An added complication in confusion estimates is {\em clustering} 
of sources which increases the confusion limits with an increasing severity 
as the source counts get steeper -- \citet{takeuchi04} show how the effect can
be factors of a few or more for super-Euclidean slopes.  
An attempt to estimate the effect
quantitatively is beyond the scope of this paper, but as we have clearly
seen, we often detect close pairs or groups of likely counterparts within 
the FIR beam size, so the confusion limits cited above are likely lower limits. 
It thus should come as no surprise that the ISOPHOT FIR galaxy
counts, both ours and others in the literature, have been significantly affected
by confusion, as probably are the very deepest MIPS FIR counts.
One should also be aware that any upward bumps in source count slopes when 
approaching the confusion regime, 
are likely due to confusion noise effects \citep[e.g.]{petri01} and
not real sources, especially in the case of super-Euclidean counts such as
these.

In summary, the effects of confusion and blending on source counts, 
their likely too high normalization and differences in slopes, 
must be properly taken into account when modelling the count slopes with galaxy
populations. Luminosity functions based on FIR counts 
will be affected as well.

\section{Summary}

We have presented follow-up observations in the NGP fields of the
ISOPHOT EBL Project.  This FIR survey complemented the other major
$ISO$ FIR projects of ELAIS, FIRBACK and Lockman Hole, by detecting
sources in three bands, and also by being able to determine
the absolute level of the CIRB \citep{juvela09}. 

The NGP fields consist of 1.64 sq.deg area and 25 unique 
FIR sources in the 90 $\umu$m band and 22 in the 150 and 180 $\umu$m.
We imaged these fields with the NOT in the optical and NIR, and
employed SED fitting techniques as well as morphological analysis to
determine the counterparts of the FIR sources in the 1\arcmin\ 
radius ISOPHOT error circles.

Only five sources were securely identified with a single bright nearby
galaxy.  One of these is a local dwarf galaxy, but all the others are
IR-luminous $L_{IR} \approx 10^{11}$, fairly massive 0.3-1.0$m_{\star}$ 
star-forming spirals at redshifts of $z<0.2$. Their star-formation rates
range from 7 to 26 $M_{\sun} \ yr^{-1}$, dust temperatures are $T_d \sim 20$~K
and dust masses $M_d \sim 10^8 \ M_{\sun}$. Such cold and fairly local 
galaxies have been found before in $ISO$ and $Spitzer$ FIR surveys selected
in the 100-200 $\umu$m range.  We note in addition that in our data most of
these appear to have disturbed morphologies showing signs of present or past 
interactions .

Nine more FIR targets were securely identified with {\em multiple}
galaxies.  It turns out, however, that these cases have essentially identical
physical characteristics with the first single-galaxy counterpart group,
with the slight difference that their redshift range extends to $z\sim0.3$. 
Thus 2/3 of our FIR galaxies are definitely relatively normal IR-luminous
cold star-forming galaxies at $z<0.3$.

Half of the remaining cases, nearly 20 per cent of the total, also have a 
bright star-forming galaxy in the field, but to be explained the FIR flux 
needs additional contribution from an optically fainter galaxy.  Finally, the
second half of the remaining cases have only fainter galaxies in the area.
In all these cases we were able to find two or three good candidates
for LIRGs or ULIRGS in the range $z=0.4-0.8$, which could explain the FIR flux.
Without further spectroscopy the exact
identification remains ambiguous however.

Given the large number of blends, we tested the often ignored aspect of
how the confused sources affect the FIR source counts themselves.  By using
both our results and those from previous surveys, we showed that significant
reshuffling of FIR fluxes is very likely to happen:  a large fraction of
apparent FIR sources at the 200-500 mJy range are in fact bound to be
double or triple sources each with fluxes in the range 50-300 mJy.  This
has the effect of both decreasing the normalization of the FIR counts by
a factor of $\sim 1.5-2$ and also steepening the count slope somewhat, 
especially at the faint end.  This has to be taken into account when
modelling FIR source counts.

Finally, many of the NGP ISOPHOT FIR sources appear to be part of significant
galaxy concentrations, pairs, groups or perhaps even clusters. What
this tells about the FIR sources and the reasons why they are 
IR-luminous, will be investigated in more detail in another paper.

\section*{Acknowledgments}

We thank the referee for valuable comments which improved the paper. 
JKK acknowledges financial support from the Academy of Finland (project 8121122)
and MJ acknowledges the support of the Academy of Finland grants no. 
105623, 124620, and 115056. This work is 
based on observations made with the Nordic Optical Telescope, operated on the 
island of La Palma jointly by Denmark, Finland, Iceland, Norway, and Sweden, 
in the Spanish Observatorio del Roque de los Muchachos of the Instituto de 
Astrofisica de Canarias and on observations with the Infrared Space Observatory
ISO. ISO is an ESA project with instruments funded by ESA member states 
(especially the PI countries France, Germany, The Netherlands, and the UK) 
and with participation of ISAS and NASA.
This work has made use of the SDSS, 2MASS and NED archives.
Funding for the SDSS and SDSS-II has been provided by the Alfred P. Sloan 
Foundation and other Participating Institutions.
The Two Micron All Sky Survey, is a joint project of the University of 
Massachusetts and the Infrared Processing and Analysis Center/California 
Institute of Technology, funded by the National Aeronautics and Space 
Administration and the National Science Foundation.
The NASA/IPAC Extragalactic Database (NED) is operated by the Jet Propulsion 
Laboratory, California Institute of Technology, under contract with the 
National Aeronautics and Space Administration.

\appendix

\section{Discussion of individual ISOPHOT fields}

\subsection{Bright unambiguous galaxies}

{\bf NGP01} The bright ($R$=16.59, $K$=13.77) counterpart is identified as
2MASSXJ13412738+4042166 with a redshift of $z=0.088$. The optical data show
a disk dominated 
galaxy with wide arms, typical
to late spirals, bar structure identified with GALFIT 
and a somewhat disturbed morphology at
the southern edge, typical to e.g.\ merger remnants.  
HYPERZ fits perfectly a fully evolved 
(nominally 15 Gyr old) GRASIL Sc template, the ERR03 models fit a 
cirrus-dominated SED, while the SDSS spectrum 
shows strong H$\alpha$ emission.
The FIR points alone -- detections at 150 and 170 $\mu$m, but only an upper
limit at 90 $\mu$m -- suggest a strong cool cirrus component. We also note
that there is another bright star-forming galaxy at exactly the same redshift 
just outside the FIR detection circle, 2.5 arcmin (or 250 kpc projected
distance) away from the adopted counterpart.

{\bf NGP03} The bright ($R$=17.12, $K$=14.20) counterpart is identified as
2MASSXJ13422216+4022017 at $z=0.131$. It has two or three 
fainter galaxies clearly in close interaction.  
GALFIT modelling shows a large bulge-dominated disk of 6.9 kpc scale length 
with wide spiral arms and the subtraction of bulge and disk components reveals 
a ring-like structure extending towards two of the companions, and possibly 
bar-like structure at the nucleus.  Again, the SDSS spectrum shows strong
star formation.
The FIR SED is now warmer (detected at 90 $\mu$m but not at 
180 $\mu$m) and our best fitting SED template is, indeed, that of a prototype 
interacting starburst galaxy NGC 6090, further corroborating its starburst 
nature.  Note that this template has more power in the cooler FIR range  
than an M82 SED would have.  The ERR03 model is similar to NGP01.

{\bf NGP20} The bright ($R$=17.75, $K$=14.80) galaxy in the detection 
circle has no SDSS spectra, but it does have a strong HYPERZ fit 
($\chi^2 = 0.2$ at $z=0.137$)
with a fully evolved Sc template, consistent with the ERR03 fit as well.  
It has an undisturbed disk and a small but bright bulge in the optical.
There is another bright galaxy of early type just outside the field 
1.1 arcmin away (2MASX J13494050+3907555) which has a known redshift of 
$z=0.143$ 
making the photometric redshift of the first galaxy, the
adopted counterpart, very plausible.

{\bf NGP22} The only bright galaxy ($R$=17.70, $K$=14.52) in the vicinity 
is the edge-on disk galaxy SDSSJ135054.71+385847.2 at a redshift of 
$z=0.086$.  The disk is somewhat warped, though there are no obvious major
partners nearby. The SDSS spectrum shows $H\alpha$, though not much else, 
and the overall SED fits reasonably the evolved Sc template at that redshift,
though the optical/NIR SED would be better fit with an M82 template.
This galaxy is 2-3 times less massive than the previous four
objects, and the SFR predicted by both GRASIL and ERR03 is also smaller, 
but still a little higher than ordinary quiescent galaxies. ERR03 again fits
a cirrus dominated galaxy.

{\bf NGP24} The counterpart galaxy of this FIR source is the most nearby 
galaxy in our sample, UGC 08793 at a redshift of $z=0.0081$ and it extends
over 1.5\arcmin\ on the sky.  Its IR luminosity is more than 
2 orders of magnitude less than the other bright sample galaxies above. 
The SDSS spectrum of the center shows a typical late type
spiral spectrum, NED classifies it as Sd, and our overall SED shape 
is well fit with a template of local Scd dwarf galaxy NGC 6946. 
This is the only case where ERR03 suggests a pure cirrus spectrum.
It has numerous bright HII regions, several small
satellite galaxies, or giant HII regions, and the disk appears slightly 
warped.

\subsection{Confused fields}

{\bf NGP02}: There is a significant concentration of $K\sim15-17$ mag 
galaxies within and just outside the ISOPHOT area -- there are four galaxies
with $K<16$ mag and a further four with $K<17$ mag. None have spectroscopy
available and none of the galaxies within the detection circle produce 
satisfactory SED fits when ISOPHOT data is included.  
However, the four brightest galaxies all have optical/NIR 
photometric redshifts of $z\approx0.28$, along with further three fainter ones.
They include a tidally disrupted interacting pair, two Sb-type SEDs, as well as
two bright early type galaxies.  A 
combination of two or three of the disk galaxies at this
redshift is well able to produce the required FIR flux.
An exact identification of NGP02 is thus ambiguous, 
though a mixed origin is very likely.

{\bf NGP04}:  There are no bright ($K<16$, $R<20$) galaxies in this field.  
The FIR detection is 130\arcsec\ 
away from an X-ray detected galaxy cluster RX J1342.8+4028 at $z=0.699$.  
The HYPERZ fits to SDSS and RIK data find a large concentration of
$z\sim0.6$ galaxies in the ISOPHOT area, which most likely belong
to the cluster.  Specifically, there are three red galaxies 
within the detection circle which fit quite well an Arp 220 SED at 
those redshifts -- and the fits are even better if the FIR is a 
sum of two of them. Two of these sources, including the brightest object 
in the field, are clearly disturbed objects. A sum of faint galaxies is
thus the likeliest counterpart for the ISOPHOT detection.

{\bf NGP05}: The brightest galaxy is a morphologically disturbed barred late 
type spiral; there are no spectra available. 
The galaxy fits reasonably an evolved Sc template at $z\approx0.25$, 
though it would need $\sim2$ times more FIR flux to fit well the FIR points. 
Other galaxies may thus contribute to the FIR flux; 
and indeed there are 2 other galaxies ($K<16.6$ mag) with photometric 
redshifts at $z\sim0.2$, best-fit SEDs (when fit is done excluding the 
ISOPHOT points) of early and late type spirals and
disk morphologies. A confused counterpart is thus likely.

{\bf NGP06}: The brightest galaxy is 2MASSX J13430669+4014314 with an SDSS
redshift of $z=0.163$. 
Both the optical spectrum and optical/NIR SED indicate
an Sb-type, but the SED including all three FIR points 
does not fit well any of our templates at 
that redshift -- the NGP 6090 template gives a reasonable fit at the
$2\sigma$ error levels of the FIR fluxes, but the optical
spectrum is not that of a starburst.
It is possible that the FIR detection come from 
some combination of a warm (contributing to the 90 $\mu$m 
band) and a cold object (the 180 $\mu$m flux).
There are also
two fainter interacting pairs in the detection circle
which both would reasonably fit an Arp 220 template at $z\sim0.5-0.6$, 
contributing to the 90 and 150 $\mu$m fluxes in that case.
The identification thus remains ambiguous.

{\bf NGP07}:  
\citet{juvela00} list this object 
with two positions in the 90 $\mu$m catalogue (NGP 07 and 08, 
separated by 20 arcsec) and a single detection in 150 and 
180 $\mu$m, suggesting
a complex origin for the FIR flux. There are 4 fairly bright galaxies 
in the area, though only one $K<16$ mag. By far the brightest appears to be 
an early-type spiral or lenticular 
while the three others are also all early type disks.  
None of these alone fit the SED templates over all the 
wavelength range. A confused FIR source, in between
$z\sim0.25-0.40$ based on photometric redshifts, is highly probable.

{\bf NGP09}:  
The two brightest galaxies in the region, 
only one is $K<16$ mag, are disk galaxies, but we
cannot fit well either of them and the FIR flux is too strong to be 
associated even with their sum.  There are, however, several
very red galaxies giving acceptable fits to ULIRG templates in a wide
redshift range of $z\approx0.2-0.8$.
We also note that this is one of three sources in the sample of 22 NGP FIR 
targets which do not have quality flag values of $q\geq3$ in any FIR band.

{\bf NGP10}:  There are no galaxies brighter than $K=16$ mag 
in the field and no object provides an acceptable HYPERZ fit. 
There are, however, half a dozen galaxies 
redder than $R-K>4$ in the area including two EROs, which could contribute
in combinations at $z\approx0.5-1.0$ where their optical/NIR photometric
redshifts place them.

{\bf NGP11}:  No single galaxy fits well the template SEDs, though  
there are two $K\approx15.5$ disk galaxies in the field, the sum of
which can explain the FIR points if both are Sc-types at their
photometric redshift of $z\sim0.3$.  They 
belong to an apparent concentration of galaxies partially inside the ISOPHOT
area.  We find, however, another fainter
red disturbed galaxy which could contribute as a higher redshift ULIRG.

{\bf NGP12}: 
Four galaxies, all $K\sim16$ mag or brighter, produce excellent exponential 
profiles with GALFIT and good optical/NIR SED fits to spiral or starburst 
templates in the
range $z=0.10-0.19$. One, or even two, of them appear to be interacting 
with another
galaxy. However, none of them fit our full SED templates alone, though a sum 
of them could explain the required FIR flux.
To complicate matters, there are four fainter red $R-K\approx 4$ galaxies 
in the field, any of which fit reasonably well a full ULIRG template in the 
range $z=0.4-0.55$.  The very ambiguous aspect of this source is 
highlighted by the fact that
\citet{juvela00} list this target as two objects (NGP 12 and 13) separated
by 23\arcsec\ in the 90 $\mu$m catalogue, while there is only one detection
in the 150 $\mu$m list.

{\bf NGP14}
There are two bright nearby spirals in the area. The brighter one 
($R$=17.32, $K$=14.19) 
is 2MASSX J13473443+3931515, with an SDSS redshift of $z=0.16$. The optical
spectrum
looks to be of a typical star-forming galaxy, and the GRASIL interacting 
starburst NGC 6090 template gives a perfect match to the overall SED. 
The optical image reveals a disk galaxy
with a large asymmetric arm.
This case would fall into the bright unambiguous class were it not for the
second spiral ($K$=15.22) which also fits very well our templates, an evolved
Sc-galaxy at a photometric $z=0.08$.  Both galaxies must contribute to the
FIR flux.

{\bf NGP15}:  
This is one of the most ambiguous cases. There is a tight group of 5--7
galaxies (all fainter than $K\approx15.5$ mag), 
including two interacting pairs, 25\arcsec\ from the FIR location. 
Two or three of these galaxies together would fit reasonably the 
FIR points with starburst templates at $z\approx0.3$.
Additionally, however,
there are three very red sources 
elsewhere inside the FIR area. If their elliptical-like optical/NIR colours are
combined with the FIR flux, they fit well ULIRG templates at 
$z\sim0.6-0.9$.  And, in fact, these sources belong to a distribution
of the largest concentration of EROs found in this follow-up survey,
mostly lying just outside the ISO detection circle. 
NGP15 is one of the three sources with lower quality flag values of $q<3$.

{\bf NGP16}:  
There are two separate 90 $\mu$m detections just 22\arcsec\
from each other (NGP16 and NGP17), 
that are connected with a single detection at longer 
wavelengths.  A significant concentration of galaxies 
is seen in the area.  There are two disk galaxies in the center of the
area, with early type spirals as best-fit SEDs at photometric redshifts of
$z\sim0.2$, but even a sum of these would not yet explain the FIR fluxes.
Intriguingly, there are also three individual fainter and red ($R-K\sim4$)
galaxies in the detection area which each give a good fit to starburst or
Arp 220 SEDs at redshifts ranging $z=0.1-0.5$.
One of these is a clear interacting galaxy.
and another a FIRST radio source 
whose radio flux is consistent with the fitted Arp 220 SED. Any one of these
could be the true counterpart, unless it is not the confused sum of
the brighter galaxies, or a combination of all. 
This is a highly ambiguous case.

{\bf NGP18}:  This field is within a galaxy cluster ZwCl 1346.9+3931.
Three bright disturbed spirals are found in the ISOPHOT area, and several 
more just outside, all with spectroscopic SDSS redshifts of $z=0.13-0.14$. 
The brightest ($R$=17.02, $K$=14.19) of the spirals, 2MASX J13490845+3917219, 
is a star forming galaxy with Seyfert-like emission lines, 
and the overall SED actually 
fits well an evolved Sc template. On the other hand, since the other 
spirals in the area must also contribute to the FIR flux (optical/NIR SEDs 
are fit with normal spiral and even starburst SEDs), we chose to 
include NGP18 in
the confused sample, rather than the bright unambiguous sample. 
It is clear though that the FIR counterpart(s) lies at 
the redshift of $z\approx0.13$.

{\bf NGP19}: The ISOPHOT circle does not have very bright galaxies, though
there are many at both sides of $K\approx16$ mag. The 90 $\mu$m flux
is the strongest of the whole sample, making this our only source with a 
clearly declining SED longwards of 100 $\mu$m.
Morphologically, according to GALFIT, two of the brighter ones
are face-on disks with stellar-PSF and/or strong bulge components. 
Both fit well an Arp 220 template at $z\approx0.35$, though without spectra 
it is impossible to speculate more about 
starbursts or obscured AGN, or to decide between them.  
Contributions from other galaxies cannot
be ruled out either: the second brightest galaxy in the field is a  
warped interacting disk galaxy, it does not fit any template by itself, and 
there are three fainter very red galaxies in the area giving
reasonable fits to ULIRG or ERO templates in redshift ranges of 
$z\approx0.3-0.6$.

{\bf NGP21}:  
There are no galaxies brighter than $K=16$ (the bright object seen in 
Fig.~\ref{FCs-medfaint1} is a star), and none of the
sources give good SED fits. The brightest source is a disk galaxy, but no
template fits the FIR points. There are also three very faint EROs in the field
which could conceivably contribute if they have ULIRG type SEDs, though
none of them alone.
This FIR source is the third ISO target with a lower quality flag value 
of $q<3$.

{\bf NGP23}:  This field again holds a significant concentration of 
bright galaxies.  The brightest 
($R$=16.73, $K$=13.60) galaxy has the GALFIT profile of an elliptical or
strongly bulge dominated disk galaxy and it fits reasonably an overall 
Sb-type SED.  The two next brightest galaxies have SDSS spectra ($z\approx0.20$)
showing star formation, and they are disks, as are two other bright galaxies, 
all with optical/NIR spectral shapes consistent with spirals.
Most of these galaxies {\em must} contribute toward the total FIR flux,
and the identification thus remains ambiguous.

{\bf NGP25}: This case is somewhat ambiguous though there is a single bright
galaxy ($R$=16.65, $K$=13.61) in the ISOPHOT area: it 
appears to be an early type galaxy (Sa template fits the optical/NIR SED), 
with some evidence of structure beneath a GALFIT de Vaucoulers 
profile.  It does not have a spectrum, but there is a galaxy of the same size 
and brightness 47\arcsec away to the NE just outside the detection circle. Its
SDSS redshift is $z=0.118$ and it has a clear disk-like GALFIT profile, and
the best fit SED is an Sb-type. Contributions from (at least) both of these
galaxies would be needed, however, to explain the FIR flux. 
An Sc-type SED 
could fit the overall SED alone 
but the morphology of neither galaxy favours this option.

\begin{landscape}

   \begin{table}

      \small
      \caption[]{Photometry of the unambiguous bright counterparts from 
our $R$, $I$, $K$-band NOT imaging, our ISOPHOT observations (Juvela 2000),
while the $ugriz$ data are from the SDSS archive. Spectroscopic SDSS redshifts 
are given with 3 decimal places, whereas our own Hyperz-photometric redshifts 
with two decimal places. Note that the RA and DEC refer to the optical/NIR 
position of the FIR counterpart, {\em not} the FIR detection.
}
         \label{table-photo-1}
         \begin{tabular}{lllccccccccllll}
            \hline
            \noalign{\smallskip}
               & RA  & DEC & $u$ & $g$ &  $r$ & $R$ &  $i$  &  $I$ &  $z$ &
            $K$  &  $90 \umu$m  & $150 \umu$m  & $180 \umu$m  & redshift \\
            \noalign{\smallskip}
            \hline
            \noalign{\smallskip}
NGP01 & 13h41m27.4s & $40^{\circ}42\arcmin16\arcsec$ & 18.80$\pm$0.07 & 17.46$\pm$0.02 & 16.80$\pm$0.03 & 16.59$\pm$0.01 & 16.41$\pm$0.02 & 16.03$\pm$0.01 & 16.24$\pm$0.10 & 13.74$\pm$0.01 & $<170$  &   $490\pm60$  &   $500\pm60$  & 0.088 \\  
NGP03 & 13h42m22.2s & $40^{\circ}22\arcmin02\arcsec$ & 19.44$\pm$0.09 & 18.11$\pm$0.01 & 17.34$\pm$0.01 & 17.10$\pm$0.01 & 16.93$\pm$0.01 & 16.47$\pm$0.01 & 16.69$\pm$0.03 & 14.17$\pm$0.01 & 170$\pm$50  &   $200\pm70$  &   $<300$  & 0.131 \\  
NGP20 & 13h49m34.9s & $39^{\circ}07\arcmin30\arcsec$ & 20.09$\pm$0.14 & 18.70$\pm$0.02 & 17.93$\pm$0.01 & 17.75$\pm$0.01 & 17.53$\pm$0.02 & 17.13$\pm$0.01 & 17.28$\pm$0.02 & 14.77$\pm$0.01 & $<150$  &   $150\pm60$  &   $200\pm80$  & 0.14 \\  
NGP22 & 13h50m54.7s & $38^{\circ}58\arcmin47\arcsec$ & 19.84$\pm$0.18 & 18.60$\pm$0.04 & 17.84$\pm$0.02 & 17.70$\pm$0.01 & 17.45$\pm$0.02 & 17.06$\pm$0.01 & 17.25$\pm$0.08 & 14.49$\pm$0.01 & $<120$  &   $320\pm110$  &   $420\pm110$  & 0.086 \\  
NGP24 & 13h52m35.0s & $38^{\circ}42\arcmin18\arcsec$ & -- &  -- &  -- &  18.11$\pm$0.01 & -- & 18.04$\pm$0.02 & -- & 15.32$\pm$0.05 &  $<130$  &   $260\pm60$  &   $280\pm80$  & 0.008 \\  
            \noalign{\smallskip}
            \hline
         \end{tabular}
   \end{table}

\end{landscape}

\begin{landscape}
   \begin{table}
      \vspace{3cm}
      \caption[]{A sample of
photometry from all the NOT data in the NGP fields as described
in Sections~2.2 and~3.1 of the Paper is shown here. The full catalog
of 15714 sources is {\em available electronically 
from CDS (http://cdsarc.u-strasbg.fr/)} -- contact the first author for access
before the paper appears in print.
Note that the catalog contains sources also from outside the ISOPHOT error 
circles.
The columns list the ALFOSC field {\em id},
a running object number N within that field, RA and DEC, then the SExtractor
CLASS ($CL$) galaxy/star classification parameter from ALFOSC data,
$R$, $I$ and $K$ SExtractor total magnitudes (e.g.\ $R_{t}$),  
2.7\arcsec\ aperture magnitudes (e.g.\ $R_{2.7}$), and their errors 
(e.g.\ e$R_{t}$), the SDSS DR7 
$ugriz$ magnitudes and their errors, and finally
the {\em IP} indicates whether the object is within 60\arcsec\ of a given 
numbered ISOPHOT FIR NGP source.  
}
         \label{table-photo-2}
            \tiny
         \begin{tabular}{rrcccrrrrrrrrrrrrrrrrrrrrrrl}
            \hline
            \noalign{\smallskip}
            $id$ &  N  & RA  &  DEC &  $CL$  & $R_{t}$  & $R_{2.7}$  &   $I_{t}$  & $I_{2.7}$  &  e$R_{t}$  & e$R_{2.7}$ &   e$I_{t}$  & e$I_{2.7}$  &   $K_{t}$  & $K_{2.7}$  & e$K_{t}$  & e$K_{2.7}$  &  $u$  &  $g$  & $r$  & $i$  & e$z$  &  e$u$  &  e$g$  & e$r$  & e$i$  & e$z$ & {\em IP} \\
            \noalign{\smallskip}
            \hline
            \noalign{\smallskip}
 1  &347 & 205.340103 &  40.733978 & 0.77 &  22.84 &  22.85 &  23.09 &  22.93 &   0.06 &   0.06  &  0.13  &  0.11  &  0.00 &   0.00 &   0.00 &   0.00 &  23.98 &  24.20 &  22.62 &  22.26 &  21.39 &   0.93 &   0.40  &  0.17  &  0.20 &   0.48  &  0 \\
  1 & 348&  205.340302&   40.722130&  0.37&   24.95&   24.87&   25.69&   24.67&    0.22&    0.36 &   0.74 &   0.53 &   0.00&    0.00&    0.00&    0.00&    0.00&    0.00&    0.00&    0.00&    0.00&    0.00&    0.00 &   0.00 &   0.00&    0.00 &   0 \\
  1 & 349&  205.340500&   40.694592&  0.41&   23.33&   23.35&   22.26&   22.30&    0.10&    0.09 &   0.07 &   0.06 &   0.00&    0.00&    0.00&    0.00&    0.00&    0.00&    0.00&    0.00&    0.00&    0.00&    0.00 &   0.00 &   0.00&    0.00 &   1 \\
  1 & 350&  205.340698&   40.700760&  0.75&   23.50&   23.50&   22.15&   22.11&    0.10&    0.10 &   0.05 &   0.05 &  18.80&   18.91&    0.13&    0.19&    0.00&    0.00&    0.00&    0.00&    0.00&    0.00&    0.00 &   0.00 &   0.00&    0.00 &   1 \\
  1 & 351&  205.340805&   40.696388&  0.03&   19.47&   19.96&   19.09&   19.55&    0.01&    0.01 &   0.01 &   0.01 &  17.32&   17.64&    0.09&    0.06&   20.49&   20.01&   19.55&   19.66&   19.66&    0.15&    0.03 &   0.03 &   0.07&    0.33 &   1 \\
  1 & 352&  205.340805&   40.739101&  0.01&   20.31&   21.51&   19.87&   21.08&    0.03&    0.02 &   0.03 &   0.02 &   0.00&    0.00&    0.00&    0.00&   24.04&   21.36&   20.78&   20.61&   20.87&    1.67&    0.07 &   0.06 &   0.09&    0.59 &   0 \\
  1 & 353&  205.341095&   40.697529&  0.16&   22.15&   22.76&   21.47&   21.92&    0.07&    0.05 &   0.06 &   0.04 &  19.18&   19.33&    0.26&    0.28&   24.13&   23.66&   23.37&   22.14&   22.53&    0.98&    0.28 &   0.31 &   0.19&    0.94 &   1 \\
  1 & 354&  205.341797&   40.660759&  0.65&   22.48&   23.20&   22.56&   23.18&    0.07&    0.07 &   0.13 &   0.11 &   0.00&    0.00&    0.00&    0.00&    0.00&    0.00&    0.00&    0.00&    0.00&    0.00&    0.00 &   0.00 &   0.00&    0.00 &   0 \\
  1 & 355&  205.341995&   40.667881&  0.03&   19.70&   20.12&   18.96&   19.32&    0.01&    0.01 &   0.01 &   0.00 &  16.01&   16.25&    0.03&    0.02&   21.70&   21.18&   19.97&   19.40&   19.23&    0.44&    0.11 &   0.06 &   0.05&    0.17 &   0 \\
  1 & 356&  205.341995&   40.683739&  0.01&   22.11&   22.43&   21.25&   21.51&    0.05&    0.04 &   0.04 &   0.03 &  18.64&   18.98&    0.22&    0.20&   24.14&   22.91&   21.87&   21.69&   20.83&    1.72&    0.26 &   0.16 &   0.23&    0.56 &   1 \\
  1 & 357&  205.342407&   40.657299&  0.44&   23.54&   23.68&   23.40&   23.71&    0.13&    0.13 &   0.20 &   0.23 &   0.00&    0.00&    0.00&    0.00&    0.00&    0.00&    0.00&    0.00&    0.00&    0.00&    0.00 &   0.00 &   0.00&    0.00 &   0 \\
  1 & 358&  205.342499&   40.680012&  0.51&   24.56&   24.47&   23.58&   23.67&    0.21&    0.25 &   0.15 &   0.21 &   0.00&    0.00&    0.00&    0.00&    0.00&    0.00&    0.00&    0.00&    0.00&    0.00&    0.00 &   0.00 &   0.00&    0.00 &   0 \\
  1 & 359&  205.342896&   40.730228&  0.04&   19.24&   19.66&   18.55&   18.91&    0.01&    0.00 &   0.00 &   0.00 &  15.69&   15.99&    0.02&    0.01&   26.06&   20.85&   19.41&   18.93&   18.43&    1.61&    0.07 &   0.03 &   0.03&    0.11 &   0 \\
  1 & 360&  205.342896&   40.728432&  0.38&   24.33&   24.19&   23.70&   23.53&    0.23&    0.20 &   0.22 &   0.19 &   0.00&    0.00&    0.00&    0.00&    0.00&    0.00&    0.00&    0.00&    0.00&    0.00&    0.00 &   0.00 &   0.00&    0.00 &   0 \\
  1 & 361&  205.343002&   40.744381&  0.55&   21.68&   23.20&   21.30&   22.50&    0.06&    0.09 &   0.08 &   0.08 &   0.00&    0.00&    0.00&    0.00&   22.35&   20.94&   20.60&   20.87&   21.06&    1.58&    0.15 &   0.17 &   0.38&    2.29 &   0 \\
  1 & 362&  205.343307&   40.661411&  0.03&   18.21&   18.77&   17.55&   18.06&    0.00&    0.00 &   0.00 &   0.00 &   0.00&    0.00&    0.00&    0.00&   20.78&   19.37&   18.37&   17.90&   17.71&    0.22&    0.02 &   0.02 &   0.02&    0.05 &   0 \\
  1 & 363&  205.343399&   40.666248&  0.36&   23.06&   23.12&   22.42&   22.48&    0.08&    0.07 &   0.08 &   0.07 &   0.00&    0.00&    0.00&    0.00&    0.00&    0.00&    0.00&    0.00&    0.00&    0.00&    0.00 &   0.00 &   0.00&    0.00 &   0 \\
  1 & 364&  205.343399&   40.688320&  0.42&   24.91&   24.69&   24.61&   24.39&    0.28&    0.30 &   0.37 &   0.40 &   0.00&    0.00&    0.00&    0.00&    0.00&    0.00&    0.00&    0.00&    0.00&    0.00&    0.00 &   0.00 &   0.00&    0.00 &   1 \\
  1 & 365&  205.343597&   40.724831&  0.04&   22.02&   22.20&   21.06&   21.23&    0.04&    0.03 &   0.03 &   0.02 &  17.90&   17.86&    0.07&    0.07&   22.69&   22.98&   22.34&   21.42&   21.30&    0.40&    0.17 &   0.15 &   0.11&    0.50 &   0 \\
  1 & 366&  205.343800&   40.712950&  0.00&    0.00&    0.00&    0.00&    0.00&    0.00&    0.00 &   0.00 &   0.00 &  19.48&   19.68&    0.34&    0.37&    0.00&    0.00&    0.00&    0.00&    0.00&    0.00&    0.00 &   0.00 &   0.00&    0.00 &   1 \\
  1 & 367&  205.343903&   40.697868&  0.03&   19.24&   19.67&   18.61&   18.98&    0.01&    0.00 &   0.00 &   0.00 &  16.10&   16.28&    0.02&    0.02&   21.16&   20.28&   19.38&   19.01&   18.71&    0.27&    0.04 &   0.03 &   0.03&    0.13 &   1 \\
  1 & 368&  205.343994&   40.654282&  0.38&   24.00&   24.07&   23.81&   23.79&    0.22&    0.19 &   0.31 &   0.25 &   0.00&    0.00&    0.00&    0.00&    0.00&    0.00&    0.00&    0.00&    0.00&    0.00&    0.00 &   0.00 &   0.00&    0.00 &   0 \\
  1 & 369&  205.343994&   40.729069&  0.44&   23.64&   23.91&   23.41&   23.62&    0.15&    0.15 &   0.21 &   0.20 &   0.00&    0.00&    0.00&    0.00&    0.00&    0.00&    0.00&    0.00&    0.00&    0.00&    0.00 &   0.00 &   0.00&    0.00 &   0 \\
  1 & 370&  205.344498&   40.696041&  0.94&   22.70&   22.65&   22.16&   22.11&    0.05&    0.05 &   0.05 &   0.05 &   0.00&    0.00&    0.00&    0.00&   24.27&   23.33&   22.29&   22.05&   22.25&    0.87&    0.19 &   0.11 &   0.14&    0.72 &   1 \\
  1 & 371&  205.344604&   40.659061&  0.27&   22.01&   22.21&   21.59&   21.67&    0.05&    0.03 &   0.06 &   0.03 &   0.00&    0.00&    0.00&    0.00&   25.38&   23.29&   22.23&   21.95&   21.39&    0.99&    0.28 &   0.15 &   0.18&    0.40 &   0 \\
  1 & 372&  205.344894&   40.692242&  0.44&   24.71&   24.65&   23.91&   24.06&    0.23&    0.29 &   0.19 &   0.30 &   0.00&    0.00&    0.00&    0.00&    0.00&    0.00&    0.00&    0.00&    0.00&    0.00&    0.00 &   0.00 &   0.00&    0.00 &   1 \\
  1 & 373&  205.345093&   40.683819&  0.64&   23.72&   23.78&   22.23&   22.22&    0.13&    0.13 &   0.06 &   0.05 &  18.70&   18.83&    0.20&    0.17&    0.00&    0.00&    0.00&    0.00&    0.00&    0.00&    0.00 &   0.00 &   0.00&    0.00 &   1 \\
  1 & 374&  205.345200&   40.673759&  0.47&   23.43&   23.67&   23.13&   23.19&    0.14&    0.12 &   0.19 &   0.13 &   0.00&    0.00&    0.00&    0.00&    0.00&    0.00&    0.00&    0.00&    0.00&    0.00&    0.00 &   0.00 &   0.00&    0.00 &   0 \\
  1 & 375&  205.345200&   40.721111&  0.16&   21.87&   22.12&   21.54&   21.70&    0.04&    0.03 &   0.05 &   0.04 &   0.00&    0.00&    0.00&    0.00&   24.17&   22.64&   22.09&   22.11&   22.05&    0.93&    0.11 &   0.10 &   0.17&    0.72 &   0 \\
  1 & 376&  205.345500&   40.702600&  0.00&    0.00&    0.00&    0.00&    0.00&    0.00&    0.00 &   0.00 &   0.00 &  19.47&   19.69&    0.20&    0.39&    0.00&    0.00&    0.00&    0.00&    0.00&    0.00&    0.00 &   0.00 &   0.00&    0.00 &   1 \\
  1 & 377&  205.345596&   40.648869&  0.31&   22.76&   23.12&   22.11&   22.31&    0.11&    0.08 &   0.10 &   0.06 &   0.00&    0.00&    0.00&    0.00&    0.00&    0.00&    0.00&    0.00&    0.00&    0.00&    0.00 &   0.00 &   0.00&    0.00 &   0 \\
  1 & 378&  205.345596&   40.651150&  0.37&   24.71&   24.63&   24.64&   24.66&    0.23&    0.31 &   0.38 &   0.56 &   0.00&    0.00&    0.00&    0.00&    0.00&    0.00&    0.00&    0.00&    0.00&    0.00&    0.00 &   0.00 &   0.00&    0.00 &   0 \\
  1 & 379&  205.346405&   40.705742&  0.66&   23.41&   23.47&   23.29&   23.09&    0.12&    0.10 &   0.18 &   0.12 &   0.00&    0.00&    0.00&    0.00&    0.00&    0.00&    0.00&    0.00&    0.00&    0.00&    0.00 &   0.00 &   0.00&    0.00 &   1 \\
  1 & 380&  205.346893&   40.705860&  0.62&   23.55&   23.66&   22.56&   22.59&    0.16&    0.12 &   0.11 &   0.08 &   0.00&    0.00&    0.00&    0.00&    0.00&    0.00&    0.00&    0.00&    0.00&    0.00&    0.00 &   0.00 &   0.00&    0.00 &   1 \\
  1 & 381&  205.346893&   40.708519&  0.75&   22.72&   22.75&   22.35&   22.19&    0.07&    0.05 &   0.09 &   0.05 &  19.09&   19.06&    0.21&    0.21&   24.02&   23.88&   22.61&   21.97&   21.97&    0.92&    0.32 &   0.16 &   0.15&    0.71 &   1 \\
  1 & 382&  205.347107&   40.726608&  0.03&   19.41&   20.39&   18.82&   19.65&    0.01&    0.01 &   0.01 &   0.01 &  16.34&   16.80&    0.04&    0.03&   21.44&   20.57&   19.55&   19.22&   19.26&    0.61&    0.09 &   0.05 &   0.08&    0.40 &   0 \\
  1 & 383&  205.347198&   40.717892&  0.98&   17.64&   17.75&   16.05&   16.13&    0.00&    0.00 &   0.00 &   0.00 &  13.60&   13.63&    0.00&    0.00&   21.55&   19.40&   17.99&   16.59&   15.81&    0.20&    0.01 &   0.01 &   0.00&    0.01 &   0 \\
  1 & 384&  205.347600&   40.716520&  0.00&    0.00&    0.00&    0.00&    0.00&    0.00&    0.00 &   0.00 &   0.00 &  18.90&   18.96&    0.20&    0.19&    0.00&    0.00&    0.00&    0.00&    0.00&    0.00&    0.00 &   0.00 &   0.00&    0.00 &   0 \\
  1 & 385&  205.347900&   40.708080&  0.39&   24.24&   24.25&   23.57&   23.70&    0.21&    0.21 &   0.19 &   0.22 &   0.00&    0.00&    0.00&    0.00&    0.00&    0.00&    0.00&    0.00&    0.00&    0.00&    0.00 &   0.00 &   0.00&    0.00 &   1 \\
  1 & 386&  205.348100&   40.721030&  0.00&    0.00&    0.00&    0.00&    0.00&    0.00&    0.00 &   0.00 &   0.00 &  19.67&   19.47&    0.19&    0.30&    0.00&    0.00&    0.00&    0.00&    0.00&    0.00&    0.00 &   0.00 &   0.00&    0.00 &   0 \\
  1 & 387&  205.348206&   40.703239&  0.47&   22.35&   22.84&   21.55&   21.99&    0.08&    0.06 &   0.06 &   0.05 &   0.00&    0.00&    0.00&    0.00&    0.00&    0.00&    0.00&    0.00&    0.00&    0.00&    0.00 &   0.00 &   0.00&    0.00 &   1 \\
  1 & 388&  205.348297&   40.675339&  0.46&   24.26&   24.30&   23.55&   23.50&    0.23&    0.21 &   0.21 &   0.18 &   0.00&    0.00&    0.00&    0.00&    0.00&    0.00&    0.00&    0.00&    0.00&    0.00&    0.00 &   0.00 &   0.00&    0.00 &   0 \\
  1 & 389&  205.348404&   40.679119&  0.03&   17.57&   18.73&   17.12&   18.18&    0.00&    0.00 &   0.00 &   0.00 &  15.15&   16.04&    0.02&    0.01&   19.43&   18.25&   17.73&   17.42&   17.26&    0.10&    0.01 &   0.01 &   0.01&    0.06 &   0 \\
  1 & 390&  205.348404&   40.710121&  0.16&   22.30&   22.49&   21.39&   21.61&    0.07&    0.04 &   0.05 &   0.03 &  19.31&   19.20&    0.14&    0.24&   24.28&   23.87&   22.75&   22.13&   21.17&    0.92&    0.29 &   0.16 &   0.16&    0.36 &   1 \\
            \noalign{\smallskip}
            \hline
         \end{tabular}
   \end{table}

\end{landscape}

\label{lastpage}


\begin{thebibliography}{299}

\bibitem[\protect\citeauthoryear{Abazajian et al.}{2009}]{sdss}
Abazajian K. et al., 2009, ApJS, 182, 543

\bibitem[\protect\citeauthoryear{Bell \& De Jong}{2001}]{bell}
        Bell E.F., De Jong R.S., 2001, ApJ, 550, 212

\bibitem[\protect\citeauthoryear{Bertin \& Arnouts}{1996}]{bertin}
Bertin E., Arnouts S., 1996, A\&AS, 117, 393

\bibitem[\protect\citeauthoryear{Bolzonella et al.}{2000}]{bolzonella}
Bolzonella M., Miralles J.-M., Pell\'o R., 2000, A\&A 363, 476


\bibitem[\protect\citeauthoryear{Bruzual \& Charlot}{1993}]{bruzual}
        Bruzual G., Charlot S., 1993, ApJ, 405, 538


\bibitem[\protect\citeauthoryear{Cole et al.}{2001}]{cole01}  
        Cole S., et al., 2001, MNRAS, 326, 255 


\bibitem[\protect\citeauthoryear{Dale et al.}{2001}]{dale01}  
        Dale D.A., Helou G., Contursi A., Silbermann N.A., Kolhatkar S.,
        2001, ApJ, 549, 215


\bibitem[\protect\citeauthoryear{Dale et al.}{2005}]{dale05}  
        Dale D., et al., 2005, ApJ, 633, 857


\bibitem[\protect\citeauthoryear{Dennefeld et al.}{2005}]{dennefeld05} 
        Dennefeld M., et al., 2005, A\&A, 440, 5

\bibitem[\protect\citeauthoryear{Devlin et al.}{2009}]{devlin09} 
        Devlin M., et al., 2009, Nature, 458, 737


\bibitem[\protect\citeauthoryear{Dole et al.}{2001}]{dole01} 
        Dole H., et al., 2001, A\&A 372, 364

\bibitem[\protect\citeauthoryear{Dole et al.}{2003}]{dole03}  
        Dole H., Lagache, G., Puget J.-L., 2003, ApJ 585, 617


\bibitem[\protect\citeauthoryear{Dole et al.}{2006}]{dole06}  
        Dole H., et al., 2006, A\&A 451, 417

\bibitem[\protect\citeauthoryear{Dye et al.}{2009}]{dye09}  
       Dye S., et a., 2009, ApJ, 703, 285

\bibitem[\protect\citeauthoryear{Efstathiou \& Rowan-Robinson}{2003}]{andreas03}
        Efstathiou A., Rowan-Robinson M., 2003, MNRAS, 343, 322 (ERR03)

\bibitem[\protect\citeauthoryear{Efstathiou et al.}{2000}]{andreas00}
        Efstathiou A., Rowan-Robinson M., Siebenmorgen R., 2000, MNRAS, 313, 734
 
\bibitem[\protect\citeauthoryear{Elbaz et al.}{2002}]{elbaz02} 
        Elbaz D. et al., 2002, A\&A 384, 848


\bibitem[\protect\citeauthoryear{Farrah et al.}{2003}]{farrah03}
	Farrah D., Afonso J., Efstathiou A., Rowan-Robinson M., Fox M., 
        Clements D., 2003, MNRAS, 343, 585

\bibitem[\protect\citeauthoryear{Fernandez-Conde et al.}{2008}]{fernan08}  
        Fernandez-Conde, N., Lagache, G., Puget, J.-L., Dole, H.,
        A\&A, 481, 885

\bibitem[\protect\citeauthoryear{Fixsen et al.}{1998}]{fixsen98}  
        Fixsen D.J., Dwek E., Mather J.C., Bennett C.L., Shafer R.A., 1998,
        ApJ, 508, 123

\bibitem[\protect\citeauthoryear{Frayer et al.}{2006a}]{frayer06}  
        Frayer D.T. et al. 2006a, ApJ, 647, L9

\bibitem[\protect\citeauthoryear{Frayer et al.}{2006b}]{frayer06b}  
        Frayer D.T. et al. 2006b, AJ, 131, 250

\bibitem[\protect\citeauthoryear{Frayer et al.}{2009}]{frayer09}  
        Frayer D.T. et al. 2009, AJ, 138, 1261

\bibitem[\protect\citeauthoryear{Gil de Paz et al.}{2000}]{gildepaz}  
        Gil de Paz A., Arag\'on-Salamanca A., Gallego J., Alonso-Herrero A.,
        Zamorano J., Kauffmann G., 2000, MNRAS, 316, 357

\bibitem[\protect\citeauthoryear{Hauser et al.}{1998}]{hauser98}  
        Hauser M.G. et al., 1998, ApJ 508, 25

\bibitem[\protect\citeauthoryear{Hauser \& Dwek}{2001}]{hauser01}  
        Hauser M.G., Dwek E., 2001, ARAA 39, 249


\bibitem[\protect\citeauthoryear{Jeong et al.}{2006}]{jeong06}  
        Jeong W.-S., Pearson, C.P., Lee H.M., Pak S., Nakagawa T., 2006, 
        MNRAS, 369, 281


\bibitem[\protect\citeauthoryear{Juvela, Mattila \& Lemke}{2000}]{juvela00}  
        Juvela M., Mattila K., Lemke D., 2000, A\&A 360, 813

\bibitem[\protect\citeauthoryear{Juvela et al.}{2009}]{juvela09}  
        Juvela M., Mattila K., Lemke D., Klaas U., Leinert C., Kiss Cs.,
        2009, A\&A, 500, 763

\bibitem[\protect\citeauthoryear{Kauffmann}{2004}]{kauffmann04}  
        Kauffmann, G., et al., 2004, MNRAS, 353, 713


\bibitem[\protect\citeauthoryear{Kennicutt}{1998}]{kennicutt}  
        Kennicutt R.C., 1998, ARA\&A, 36, 189

\bibitem[\protect\citeauthoryear{Lagache et al.}{2004}]{lagache04}
        Lagache G., et al., 2004, ApJS 154, 112

\bibitem[\protect\citeauthoryear{Lagache, Puget \& Dole}{2005}]{lagache05} 
        Lagache G., Puget J.-L., Dole H., 2005, ARA\&A 43, 727

\bibitem[\protect\citeauthoryear{Malek et al.}{2009}]{malek09}
        Malek  K., et al., 2009, in proceedings 
``AKARI, a light to illuminate the misty Universe", February 16-19 2009, 
Tokyo'', arXiv:0903.3987


\bibitem[\protect\citeauthoryear{Mann et al.}{2002}]{mann02}
        Mann R.G., et al., 2002, MNRAS, 332, 549

\bibitem[\protect\citeauthoryear{Matsuura et al.}{2007}]{matsuura07}
        Matsuura S., et al., 2007, PASJ, 59, S503 


\bibitem[\protect\citeauthoryear{Oyabu et al.}{2005}]{oyabu05}  
        Oyabu S., et al., 2005, AJ, 130, 2019

\bibitem[\protect\citeauthoryear{Papovich et al.}{2004}]{papovich04}  
        Papovich C., et al., 2004, ApJS, 154, 70

\bibitem[\protect\citeauthoryear{Papovich et al.}{2007}]{papovich07}  
        Papovich C., et al., 2007, ApJ, 668, 45

\bibitem[\protect\citeauthoryear{Patris et al.}{2003}]{patris03}  
        Patris J., Dennefeld M., Lagache G., Dole H., 2003, A\&A 412, 349

\bibitem[\protect\citeauthoryear{Peng et al.}{2002}]{peng02} Peng, C.Y., 
Ho L.C., Impey C.D., Rix H.-W., 2002, AJ, 124, 266

\bibitem[\protect\citeauthoryear{Pilbratt}{2008}]{pilbratt}  
        Pilbratt G.L. 2008, SPIE, 7010, 1

\bibitem[\protect\citeauthoryear{Puget et al.}{1996}]{puget96}  
        Puget J.-L. et al., 1996, A\&A 308, L5

\bibitem[\protect\citeauthoryear{Rodighiero et al.}{2005}]{rodighiero05}  
        Rodighiero G., Fadda D., Franceschini A., Lari C., 
2005, MNRAS, 357, 449

\bibitem[\protect\citeauthoryear{Rodighiero et al.}{2006}]{rodighiero06}  
        Rodighiero G., et al., 2006, MNRAS, 371, 1891

\bibitem[\protect\citeauthoryear{Rowan-Robinson}{2009}]{rowan09}  
        Rowan-Robinson M., 2009, MNRAS, 394, 117


\bibitem[\protect\citeauthoryear{Silva et al.}{1998}]{silva98}  
        Silva L., et al., 1998, ApJ, 509, 103

\bibitem[\protect\citeauthoryear{Sajina et al.}{2003}]{sajina03}  
        Sajina A., et al., 2003, MNRAS, 343, 1365

\bibitem[\protect\citeauthoryear{Sajina et al.}{2006}]{sajina06}  
        Sajina A., Scott D., Dennefeld M., Dole H., Lacy M., Lagache G.,
        2006, MNRAS, 396, 939

\bibitem[\protect\citeauthoryear{Sanders et al.}{2003}]{sanders03} 
        Sanders D.B., Mazzarella J.M., Kim D.-C., Surace J.A., Soifer B.T.,
        2003, AJ, 126, 1607


\bibitem[\protect\citeauthoryear{Siebenmorgen \& Kr\"ugel}{1992}]{sieben92}  	
	Siebenmorgen R., Kr\"ugel E., 1992, A\&A, 259, 614

\bibitem[\protect\citeauthoryear{Soifer, Helou \& Werner}{2008}]{soifer08}  
        Soifer B.T., Helou G., Werner M., 2008, ARA\&A 46, 201

\bibitem[\protect\citeauthoryear{Stickel, Klaas \& Lemke}{2007}]{stickel07}  
        Stickel M., Klaas U., Lemke D., 2007, A\&A, 466, 831

\bibitem[\protect\citeauthoryear{Symeonidis et al.}{2008}]{symeon08}  
        Symeonidis M., Willner S.P., Rigopoulou D., Huang J.-S., 
        Fazio G.G., Jarvis M.J., 2008, MNRAS, 385, 1015

\bibitem[\protect\citeauthoryear{Symeonidis et al.}{2009}]{symeonidis09}  
        Symeonidis M., 2009, MNRAS, 397, 1728


\bibitem[\protect\citeauthoryear{Takeuchi \& Ishii}{2004}]{takeuchi04}
        Takeuchi T.T., Ishii T.T., 2004, ApJ, 604, 40


\bibitem[\protect\citeauthoryear{Taylor et al.}{2005}]{taylor05}  
        Taylor E.L., et al. 2005, MNRAS, 361, 1352

\bibitem[\protect\citeauthoryear{V\"ais\"anen et al.}{2000}]{petri00}
         V\"ais\"anen P., Tollestrup E.V., Willner S.P., Cohen M.,
         2000, ApJ, 540, 593

\bibitem[\protect\citeauthoryear{V\"ais\"anen et al.}{2001}]{petri01}
         V\"ais\"anen P., Tollestrup E.V., Fazio G.G., 
         2001, MNRAS, 325, 1241

\bibitem[\protect\citeauthoryear{Wen et al.}{2007}]{wen07}
         Wen X.-Q, Wu H., Cao C., Xia X.-Y., 2007, ChJAA, 7, 187


\bibitem[\protect\citeauthoryear{Zheng et al.}{2007}]{zheng07}
         Zheng X.Z., et al., 2007, ApJ, 670, 301


\end{thebibliography}
\end{document}